\documentclass[a4paper,fleqn,usenatbib]{mnras}

\usepackage{newtxtext,newtxmath}

\usepackage[T1]{fontenc}
\usepackage{ae,aecompl}

\usepackage{amssymb}
\usepackage{amsmath}
\usepackage{graphicx,longtable,mathrsfs,color,array}
\usepackage{epsfig,subfigure,placeins,float}
\usepackage[usenames,dvipsnames]{xcolor}
\usepackage{hyperref}

\usepackage[normalem]{ulem}
\providecommand{\adsurl}[1]{\href{#1}{ADS}}

\newcommand*{\rom}[1]{\expandafter\@slowromancap\romannumeral #1@}
\makeatother

\title[Shell Galaxies in the Illustris Simulation]{Formation and Incidence of Shell Galaxies in the Illustris Simulation}
\author[A. R. Pop et al.]{
Ana-Roxana Pop,$^{1}$\thanks{E-mail: ana-roxana.pop@cfa.harvard.edu}
Annalisa Pillepich,$^{1,2}$
Nicola C. Amorisco,$^{1,3}$
Lars Hernquist$^{1}$
\\
$^{1}$Harvard-Smithsonian Center for Astrophysics, 60 Garden St., Cambridge, MA 02138, USA \\
$^{2}$Max-Planck Institute for Astronomy, K{\"o}nigstuhl 17, 69117, Heidelberg, Germany \\
$^{3}$Max Planck Institute for Astrophysics, Karl-Schwarzschild-Strasse 1, D-85740 Garching, Germany}

\date{Accepted 2018 July 18.}

\pubyear{2018}

\begin{document}
\label{firstpage}
\pagerange{\pageref{firstpage}--\pageref{lastpage}}
\maketitle

\label{firstpage}

\begin{abstract}
Shells are low surface brightness tidal debris that appear as interleaved caustics with large opening angles, often situated on both sides of the galaxy center. 
In this paper, we study the incidence and formation processes of shell galaxies in the cosmological gravity+hydrodynamics Illustris simulation. 
We identify shells at redshift z=0 using stellar surface density maps, and we use stellar history catalogs to trace the birth, trajectory and progenitors of each individual star particle contributing to the tidal feature. 
Out of a sample of the 220 most massive galaxies in Illustris ($\mathrm{M}_{\mathrm{200crit}}>6\times10^{12}\,\mathrm{M}_{\odot}$), $18\%\pm3\%$ of the galaxies exhibit shells. 
This fraction increases with increasing mass cut: higher mass galaxies are more likely to have stellar shells.
Furthermore, the fraction of massive galaxies that exhibit shells decreases with increasing redshift.
We find that shell galaxies observed at redshift $z=0$ form preferentially through relatively major mergers ($\gtrsim$1:10 in stellar mass ratio).
Progenitors are accreted on low angular momentum orbits, in a preferred time-window between $\sim$4 and 8 Gyrs ago. 
Our study indicates that, due to dynamical friction, more massive satellites are allowed to probe a wider range of impact parameters at accretion time, while small companions need almost purely radial infall trajectories in order to produce shells.
We also find a number of special cases, as a consequence of the additional complexity introduced by the cosmological setting.
These include galaxies with multiple shell-forming progenitors, satellite-of-satellites also forming shells, or satellites that fail to produce shells due to multiple major mergers happening in quick succession. 
\end{abstract}


\section{Introduction}
\label{sec:introduction}

According to the standard hierarchical cosmological model, galaxies build up their stellar mass via star formation and successive 
mergers \citep[see, e.g.,][]{White&Rees1978, Davisetal1985, Bullock&Johnston2005, Johnston2008, Cooperetal2010, Rodriguez-Gomezetal2016}.
Often times, these merger events leave behind unique morphological features in the galaxies' extended stellar halos,
such as tidal streams, stellar shells, rings, or plumes \citep[e.g.,][]{McConnachieetal2009, MartinezDelgadoetal2010}. 
This paper focuses on a particular type of tidal debris, stellar shells, which are wide, concentric arcs of overdense stellar regions extending to large galactocentric distances.

Tidal features provide a powerful tool to study both the structure and accretion histories of galaxies \citep[e.g.,][]{Trujilloetal2008, Martinez-Delgadoetal2012, Arnaboldietal2012,  Romanowskyetal2012, Longobardietal2015, Fosteretal2014, Amoriscoetal2015}. For example, the number and distribution of shells can constrain the mass distribution of the host galaxy, as well as the timing of the merger event itself \citep{Quinn1984, Dupraz&Combes1986, Hernquist&Quinn1987a, Hernquist&Quinn1987b, Canalizoetal2007, Duc2016}. 
Several studies have proposed using 
the line of sight velocity distributions of shells to constrain the gravitational potential of the host galaxy \citep{Merrifield&Kuijken1998, Ebrova2012, Sanderson&Helmi2013}.
More generally, by studying their formation mechanisms and detailed morphologies, tidal features can be used to measure merger rates, as well as the distribution of orbital properties and mass ratios of satellites at accretion \citep{Johnstonetal2001, Johnston2008, Hendel&Johnston2015}. Future low surface brightness (LSB) surveys will soon provide both the depth and the large sample size necessary to enable these measurements. 

The first shell galaxies were reported in the Atlas of Peculiar Galaxies \citep{Arp1966}.
\cite{Malin&Carter1983} pointed out that postprocessing was helpful to reveal shell structures, 
resulting in the identification of 137 galaxies with shells. Since then,
various techniques have been developed to identify and enhance shells, including 
unsharp masking of photographic images \citep{Malin1977}, digital masking \citep{Schweizer&Ford1985}, 
model subtraction \citep{Pengetal2002}, or minimum masking \citep{Bileketal2015}.
Nonetheless, detecting shells around distant galaxies remains challenging, due to the shells' varied and sometimes irregular morphologies, as well as due to their low surface brightness levels. The majority of tidal features in early-type galaxies occur
at a surface brightness of \mbox{28 mag $\mathrm{arcsec}^{-2}$} or fainter in the V band \citep[e.g.,][]{Johnston2008, Atkinsonetal2013}, and outer shells often extend to very large galactocentric distances  (\mbox{$\gtrsim$100 kpc}), in the faint outskirts of the galaxies' stellar halos.

Early results suggest that shell-like structures are a relatively common feature of massive galaxies: 
\citet{Malin&Carter1983} found that as many as $17\%$ of isolated elliptical galaxies exhibit one or more shells. \citet{Schweizer1983} and 
\citet{Schweizer&Ford1985} estimate the incidence of shell galaxies to be around $10\%$ in early-type galaxies.
More recently, \cite{Taletal2009} report that $22\%$ of their sample of 55 nearby elliptical galaxies 
exhibit shells, and preliminary results from the MATLAS deep imaging survey indicate that 15.5\% of massive early-type galaxies have shells or streams \citep{Duc2016}. 
On the other hand, \cite{Krajnovicetal2011} only found shells in  $3.5\%$ of 260 early-type galaxies in the ATLAS$^{\mathrm{3D}}$ sample. 
Although shell-like structures have been detected around several spiral galaxies \citep[e.g.,][]{MartinezDelgadoetal2010, deBloketal2014},
shells appear to be more common in red early-type galaxies than in blue galaxies \citep[14\% against 6\%,][]{Atkinsonetal2013}, and they are therefore more commonly detected in surveys that focus on massive early-type galaxies \citep[e.g., shells are present in $45\%$ of the non-starburst galaxies in][]{RamosAlmeidaetal2011}.
Observations also hint at an environmental dependence: 48$\%$ of the shell galaxies found by \cite{Malin&Carter1983} were in isolated environments vs. $4\%$ occurring in clusters or rich groups.

Several mechanisms for the origin of shells have been proposed over the last decades. 
The most widely accepted formation scenario advocates that shells share a similar origin 
to other tidal features such as streams, tails and plumes, and that they are composed of 
debris derived from stripped satellites \citep[see, e.g., recent studies by][]{Amorisco2015, Hendel&Johnston2015}. 

Early theories invoked instead a varied range of mechanisms. For example, \citet{Fabianetal1980} suggested that shells 
correspond to regions of recent star formation in a shocked galactic wind \citep[see also][]{Bertschinger1985, Williams&Christiansen1985, Loewensteinetal1987}. 
However, observations do not indicate the presence of a significant fraction of young stars in shells, and 
this theory can not explain the interleaved shell distribution 
\citep[see, e.g.,][for more details]{Ebrova2013}.
A competing theory is based on tidal interactions \citep{Thomson&Wright1990, Thomson1991}, and it models shells as density waves induced in the cold disk by a passing galaxy.
In this weak interaction model, shells are predicted to have similar colors to the host, but evidence has accumulated for shell systems both redder or bluer than the host, and for outer shells being bluer than inner ones \citep[see, e.g.,][]{McGaugh&Bothum1990, Turnbulletal1999, Pierfederici&Rampazzo2004, Sikkemaetal2007}. Moreover, 
\cite{Wilkinsonetal2000} found no indication of a cold disk in the shell galaxy MC 0422-476, and this model also has trouble explaining the
observed strong association between kinematically distinct cores and shell galaxies \citep{Efstathiouetal1982, Forbes1992} and the occurrence of stellar shells tens of kpc away from the galaxy centers.

Extensive analytical and numerical work supports the merger scenario, in which shells correspond to overdensities of stripped stars accumulating at the apocenters of their orbits.
Early simulations of shell galaxies using test particles and restricted N-body codes \citep{Quinn1984, Dupraz&Combes1986} indicated that satellites coming in on radial orbits can reproduce the observed shell properties. \cite{Hernquist&Quinn1988, Hernquist&Quinn1989} argued that shells can also arise in more general interactions, such as companions on non-radial orbits, as well as low-mass disk interactions.
In a subsequent paper, \cite{Hernquist&Spergel1992} argued that a major merger between two disk galaxies can also produce shells. 
More recently, several analytic models have been developed in order to capture the distribution of shells for approximately radial orbits \citep[e.g.,][]{Sanderson&Bertschinger2010, Sanderson&Helmi2013}, which can also explain the physical conditions that distinguish between the formation of thin streams or wide stellar shells based on the dynamical properties of the accretion event \citep{Amorisco2015}. 

Therefore, just like streams and other low surface brightness features, shells are the outcome of the process of hierarchical assembly, and they have also been seen in the context of $\Lambda$CDM cosmological simulations \citep[see, e.g.,][]{Cooperetal2010, Cooperetal2011}.
However, most of the previous studies investigating the formation of shells have focused on idealized, isolated minor mergers, involving satellites on radial infall trajectories \citep[e.g.,][]{ Dupraz&Combes1986, Hernquist&Quinn1987b, Hernquist&Quinn1988, Hernquist&Quinn1989, Seguin&Dupraz1996, Sanderson&Bertschinger2010, Bartoskovaetal2011, Ebrova2012}. The present study aims to test the statistically dominant shell formation mechanism by using for the first time a systematic, large sample of 220 massive galaxies from the cosmological gravity+hydrodynamics Illustris simulation. 
In the context of hierarchical galaxy assembly, 
our study of redshift $z=0$ massive galaxies provides a systematic census of the morphologies of stellar shells both as a function of mass and redshift.
 In Section~\ref{sec:simulations}, we begin with a description of the Illustris simulation (\S\ref{subsec:Illustris}) and our galaxy sample (\S\ref{subsec:galaxySample}). The implementation of stellar history catalogs is detailed in Section~\ref{subsec:stellarHistoryCatalogs}, while in Section~\ref{subsec:identifyingShells} we define stellar shells similarly to observations, and we describe the two steps used to identify shell galaxies in Illustris. In Section~\ref{sec:distShells}, we investigate the mass distribution of shell galaxies, as well as the redshift evolution of the fraction of shells.
Section~\ref{sec:progenitors} is dedicated to studying the merger events that produce shells, providing an order-zero recipe for their formation (Figure~\ref{fig:threeDim}). We follow the time evolution of a representative shell galaxy in Section~\ref{subsec:mergerTree}, while special cases that depart from this recipe for shell formation are investigated in Section~\ref{sec:studyCases}.
We discuss our results in the context of stellar shell observations and galaxy assembly history in Section~\ref{sec:discussion}, and summarize our conclusions in Section~\ref{sec:conclusions}.
Appendix~\ref{sec:projectionEffects} discusses projection effects,
while Appendix~\ref{sec:logisticRegression} describes the logistic regression algorithm used to classify shell and non-shell-forming progenitors based on their mass and infall orbits.
Appendix~\ref{sec:resolution} addresses the issue of resolution for shell-forming progenitors, while Appendix~\ref{sec:complete} presents the total mass distribution and the stellar mass completeness of our galaxy sample.

\section{Simulations and Methods}
\label{sec:simulations}

\subsection{Illustris}
\label{subsec:Illustris}

This project uses galaxies from the Illustris Simulation \citep{Vogelsbergeretal2014b, Vogelsbergeretal2014a, Geneletal2014}, a cosmological hydrodynamical simulation with a periodic volume of $(106.5\;\mathrm{ Mpc})^3$, evolved from an initial redshift $z=127$ down to $z=0$ using cosmological initial conditions from WMAP-9 \citep{Hinshawetal2013}. Illustris is run using  {\small AREPO} \citep{Springel2010}  - a moving-mesh code based on a quasi-Lagrangian finite volume method, using a second-order Godunov scheme with an exact Riemann solver.
The Illustris project involves a series of realizations simulated at different resolutions. Since we are interested in identifying complex morphologies inside galaxies, we are working with the highest-resolution run (Illustris-1, which we simply refer to as Illustris), that has dark matter particles with a mass resolution of $\mathrm{m}_{DM} = 6.26 \times 10^6\, \mathrm{M}_{\odot}$, and baryons with $\mathrm{m}_{\mathrm{baryon}} \sim 1.26 \times 10^6\, \mathrm{M}_{\odot}$.

In addition to gravity+hydrodynamics, the Illustris simulations include a broad range of astrophysical processes such as metal-line cooling with radiative self-shielding corrections, chemical enrichment and stellar mass loss, stellar evolution and feedback from both supermassive black holes (SMBHs) and supernovae (SNe), AGN feedback and black hole evolution \citep[see][for more details]{Vogelsbergeretal2013, Torreyetal2014}.
The inclusion of these physical models allows Illustris to reproduce, for example, the observed evolution of the cosmic star formation rate density and galaxy stellar mass function \citep{Geneletal2014}. Illustris captures the accretion histories of galaxies with a diverse range of morphologies and colors down to $z=0$ \citep{ Snyderetal2015, Torreyetal2015}, and the current project takes advantage of this feature in order to develop a systematic study of stellar shells.

\subsection{The Galaxy Sample}
\label{subsec:galaxySample}

Our sample consists of the 220 most massive, redshift $z=0$ galaxies in Illustris. We apply a halo mass cut based on the virial mass of the host: $\mathrm{M}_{\text{200crit}}>6 \times 10^{12}\, \mathrm{M}_{\odot}$, where $\mathrm{M}_{\text{200crit}}$ is defined as the total mass inside a radius enclosing a sphere with mean density $200$ times greater than the critical density of the Universe. 
Galaxies in our sample have stellar masses $\mathrm{M}_{\mathrm{stars,tot}} \gtrsim 10^{11} \, \mathrm{M}_{\odot}$, where $\mathrm{M}_{\mathrm{stars,tot}}$ is the total mass of all star particles gravitationally bound to the halo.
The distribution of total and stellar masses for galaxies in our sample is presented in Figure~\ref{fig:completeness}. 
For the halo mass selection $>6 \times 10^{12}\, \mathrm{M}_{\odot}$, our sample is complete for galaxies with stellar masses $>3.6 \times 10^{11}\, \mathrm{M}_{\odot}$ at $z=0$.

We investigate the presence of stellar shells in the central galaxies of the selected halos.
In Illustris, halos or groups are identified using a friends-of-friends (FOF) algorithm with linking length $b=0.2$ run on the dark matter particles, with subhalos or satellites inside each halo identified using the {\small SUBFIND} algorithm \citep{Davisetal1985, Springeletal2001, Dolagetal2009}. 
We refer to central galaxies (or centrals) as the most massive galaxies in a given halo/group, i.e., galaxies that are not satellites/subhalos in the potential well of a more massive galaxy. 

\subsection{Stellar History Catalogs}
\label{subsec:stellarHistoryCatalogs}

Our goal is to identify and track the evolution of stars in shells, in order to ultimately get a better understanding of the processes leading to shell formation.
Therefore, we develop stellar history catalogs that trace the birth, trajectory and progenitors of all stars in a given $z=0$ halo. 
We define three key moments during the life of every star, relying on the binding and unbinding criteria of the adopted {\small SUBFIND} (sub)halo finder:
\begin{itemize}
\item \textit{formation time} ($t_{\mathrm{form}}$): when a gas cell (or part of its mass) is replaced by a star particle.
\item \textit{accretion time} ($t_{\mathrm{acc}}$): when the satellite to which the star belongs first enters the virial radius of the central host galaxy.
\item \textit{stripping time} ($t_{\mathrm{strip}}$): when the star stops being gravitationally bound to its parent satellite and becomes bound to the primary galaxy.
\end{itemize}

In Illustris, information about all star particles in the simulation box is stored at 136 snapshots in time, with redshifts closer to $z=0$ being sampled more frequently. Starting from the earliest snapshot (highest redshift), at each time step we match stars belonging to a given halo at $z=0$ to all stars in the simulation based on the unique ID of the star particle. When a new star is found in a given snapshot, we record its formation time. 
Next, we mark the accretion time of a star as the first time when the star particle comes to a galactocentric distance equal to or smaller than the virial radius of the host at that redshift.
Finally, we mark both the first time when a star becomes gravitationally bound to the central host (`first stripping time') and the latest time/smallest redshift when the star switches from being bound to a satellite and becomes bound to the central galaxy (`last stripping time'). For all of these different events in the life of a star, we save information about the star's galactocentric distance to the final host, relative position and velocity with respect to the central galaxy, the ID of the satellite it belongs to at the moment of formation or accretion (or the satellite it belonged to right before being stripped), as well as the gravitational potential energy the star experienced at each of these moments.

One advantage of this approach is that it allows us to easily separate stars into several key categories based on their formation and evolution. Following \cite{Pillepichetal2014}, we distinguish between:
\begin{itemize}
\item \textit{in-situ stars}: stars that form inside the primary halo potential well, or in other words, stars forming out of the gas found inside the central galaxy.
\item \textit{ex-situ stars}: also referred to as `accreted stars', ex-situ stars form inside satellites but they are stripped at a later time by the central galaxy.
\end{itemize}
Based on our definitions above, in-situ stars are those for which formation, accretion, and stripping time all coincide. In turn, ex-situ stars generally form outside the virial radius of the host (i.e., $t_{\mathrm{form}}$ precedes $t_{\mathrm{acc}}$).
A subset of the ex-situ star particles form at accretion time or later, and we distinguish these based on the fact that the ID of their parent satellite at formation time is different from the ID of the final host at the same snapshot.
In this way, ex-situ stars can be subdivided into two classes based on whether their formation time precedes or follows the time at which their progenitor satellite crosses the virial radius of the central galaxy.

Using the stellar history catalogs just described, we investigate in \S\ref{subsubsec:defineShells} whether shell material is ex-situ or in-situ, and in \S\ref{subsubsec:step2Confirm} we test whether star particles in shells share common histories (such as, e.g., same parent satellites or similar stripping times).

\subsection{Shell Galaxies Identification}
\label{subsec:identifyingShells}

In this subsection, we describe key common features through which we can distinguish stellar shells from other low surface brightness substructures. We then present our two-step approach for identifying shell galaxies in Illustris. The first step (\S\ref{subsubsec:step1Visual}) relies on visual identification based on stellar surface density maps, while the second step (\S\ref{subsubsec:step2Confirm}) involves following the history of all stars present in the halo at z=0 and identifying the progenitors responsible for forming shells.

\subsubsection{Defining Shells}
\label{subsubsec:defineShells}

\begin{figure*}
\centering
\includegraphics[width=\textwidth]{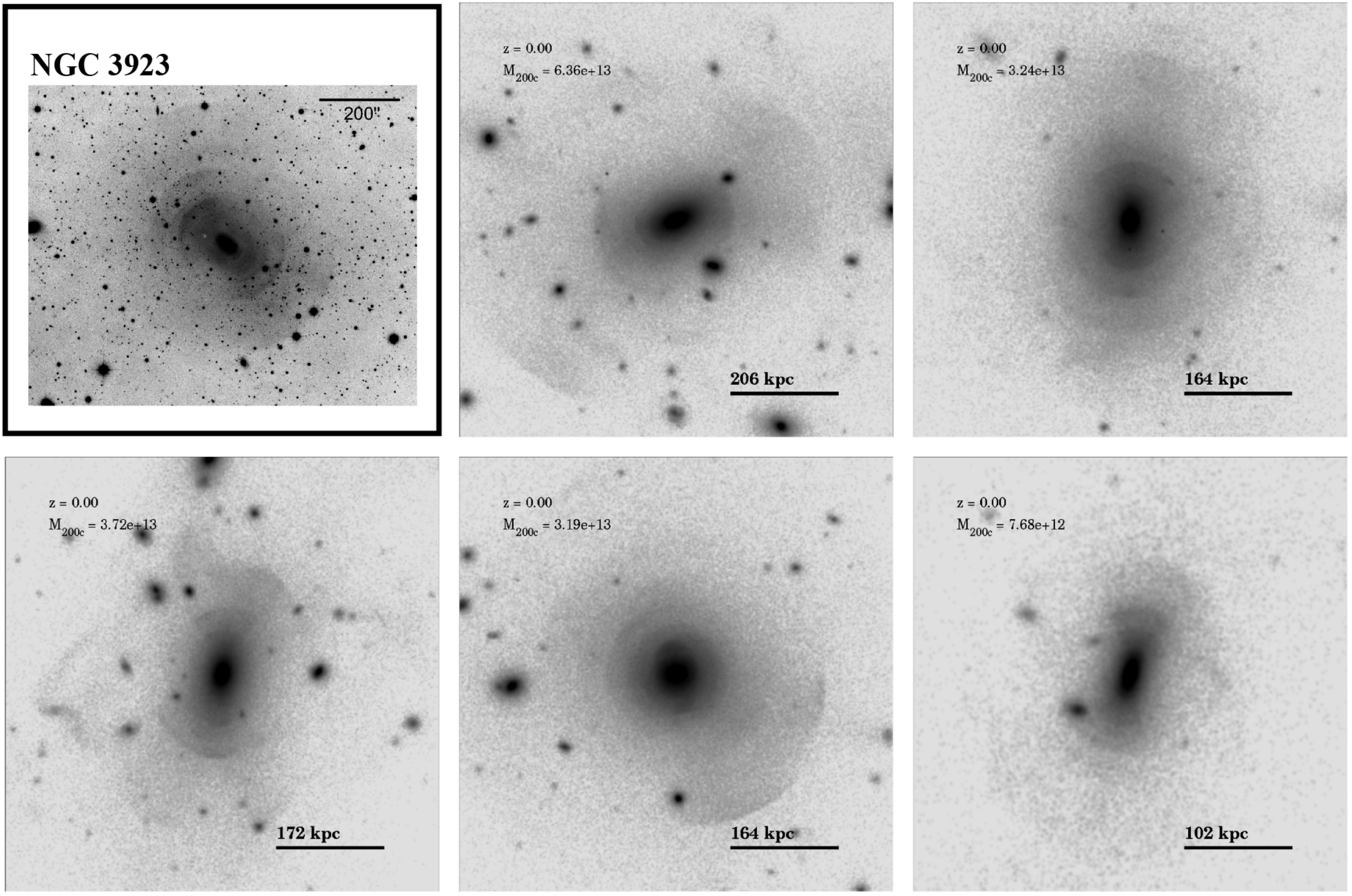}
\caption{Upper left panel shows observations of the type I shell galaxy NGC 3923, courtesy of AAO/David Malin. The other five panels show stellar surface density maps for a sample of halos from the Illustris simulation that exhibit $z=0$ stellar shells, with a wide variety of shell morphologies. The mass quoted in each figure is $\mathrm{M}_{200\mathrm{crit}}$ (in solar mass units) and the side length of each stamp corresponds to $\mathrm{R}_{200\mathrm{crit}}$.}
\label{fig:obsvsIllustris}
\end{figure*}

In our sample of simulated galaxies, we are looking for shell features similar to observations. 
Some of the most well-known shell galaxies include NGC 1316 (Fornax A) and NGC 5128 (Cen A) \citep{ Malinetal1983, Schweizer1983}. 
The observed morphologies vary significantly: the shells in NGC 1316 have large opening angles, while NGC 5128 has more narrow shells.
Systems such as NGC 3923 \citep[shown in the upper left panel of Figure~\ref{fig:obsvsIllustris}][]{Malin&Carter1983, Prieur1988, Zepfetal1995, Norrisetal2008}, NGC 1344 \mbox{\citep{Carteretal1982}}, or NGC 5982 \citep{Sikkemaetal2007} exhibit dozens of shells. However, more typical shell galaxies only have a handful of observable shells \citep{Malin&Carter1983}.

\cite{Wilkinsonetal1987} and \cite{Prieur1989} introduced a system for dividing shells into 3 classes based on their morphology. Type I shell galaxies have interleaved shells aligned along the major axis of the galaxy. Shells classified as type II are randomly distributed around the galaxy \citep[also called "all-around" systems, e.g. NGC 474,][]{Turnbulletal1999}. The third type usually encompasses those shell galaxies that cannot be classified as either type I or II, because they show irregular features or there are too few visible shells to be classified.

Figure~\ref{fig:obsvsIllustris} compares the shells in NGC 3923 (upper left corner) \citep{Malin&Carter1983} with  
stellar surface density maps of 5 shell galaxies in Illustris. 
We also find varied shell morphologies, with several simulated galaxies reproducing shell features such as those observed in NGC 1344, NGC 7600, or NGC 5128 \citep{ Carteretal1982,   Malinetal1983,Schweizer1983, Turnbulletal1999}. For example, the galaxies in the left and right bottom panels of Figure~\ref{fig:obsvsIllustris} have tidal features similar to Type I shells -- the narrow opening angles of these shells resemble those in NGC 3923. 
The galaxy in the middle bottom panel, on the other hand, would fit better in the Type II category, due to the roughly circular distribution of shells, without a clear symmetry axis.

\begin{figure*}
\centering
\includegraphics[width=\textwidth]{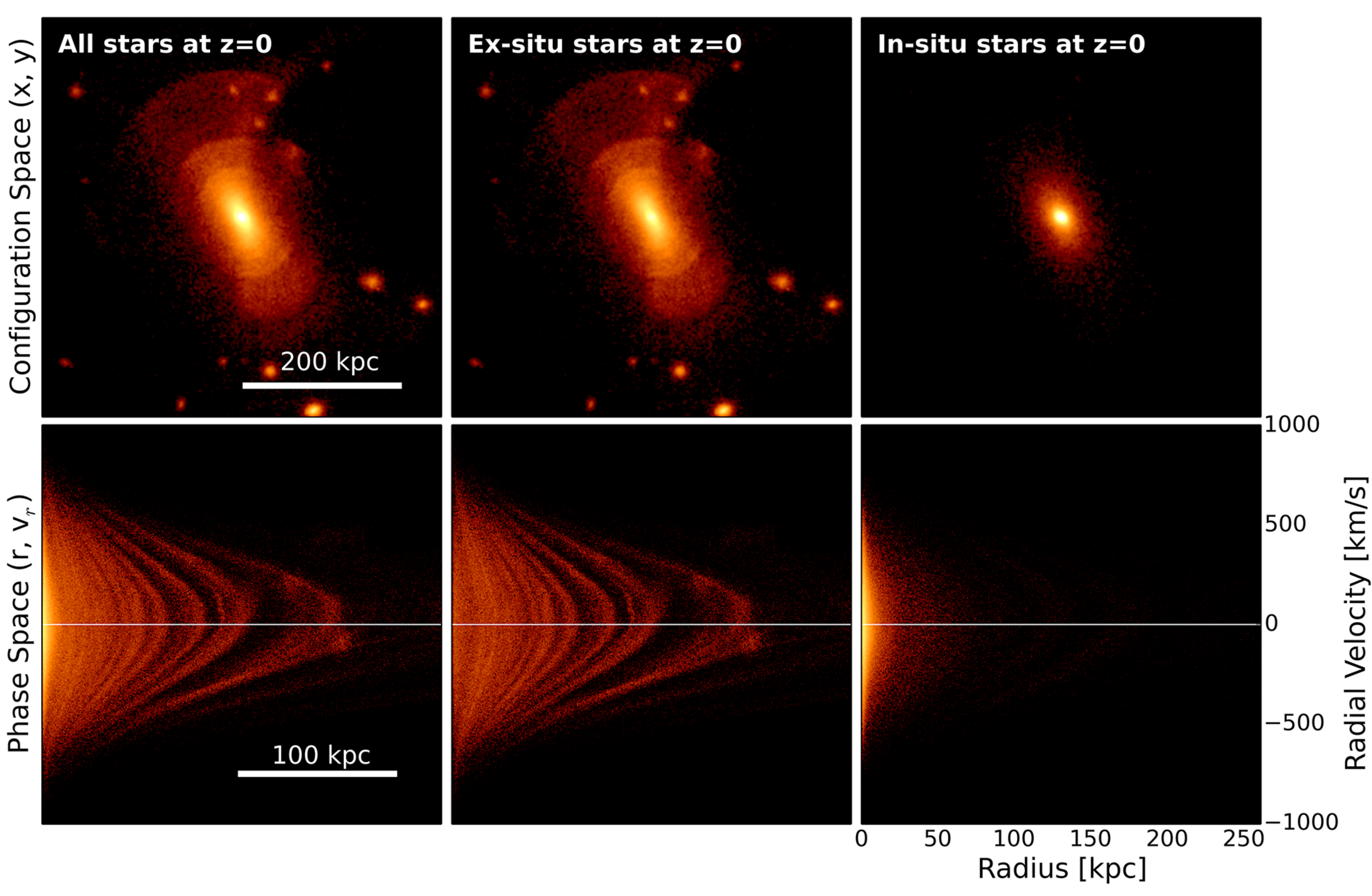}
\caption{Example of the stellar surface mass density of a shell galaxy at redshift zero (top row) and the corresponding phase-space ($v_r$ versus $r$) distribution of stars in the galaxy (bottom row). From left to right: all the stars in the halo at $z=0$ (all stars in the central galaxy and its surrounding satellites), ex-situ stars (accreted from satellites already stripped by the central), in-situ stars (formed inside the primary halo potential well). In configuration space, we identify shells as low surface brightness arc-like features interleaved on both sides of the galaxy center. In turn, shells appear as caustics in phase space, with stars spending most of their time at the apocenter of their orbits (i.e., close to the horizontal white line marking $v_r =0$). Shells are part of the ex-situ stars stripped from satellites accreted by the central galaxy at some point in the past.}
\label{fig:exsituShells}
\end{figure*}

As mentioned in the Introduction, several theories for the origin of shell substructures have been proposed over the last decades. 
The most widely accepted model is based on the idea that shells are formed through mergers, 
with stripped stars at the apocenters of their orbits forming peaks in the stellar density maps. 
In agreement with this model, and as depicted in Figure~\ref{fig:exsituShells}, shells in Illustris are composed of ex-situ stars. 
The left column of Figure~\ref{fig:exsituShells} shows all stars in the halo at redshift $z=0$, 
while the next two columns include only the ex-situ and in-situ stars, respectively. 
Thus, we identify shells in configuration space (top row of Figure~\ref{fig:exsituShells}) as overdensities of ex-situ stars located at large galactocentric distances that, differently from streams, appear interleaved on both sides of the galaxy center. Shells at greater galactocentric distances often  have larger opening angles.

Also, as exemplified by the bottom row in Figure~\ref{fig:exsituShells}, stellar shells are particularly easy to recognize in phase space, i.e. in the space of galactocentric distance vs. radial velocity ($r\,-\,v_r$), 
since they are composed of stars piling up at the apocenters of their orbits \citep[see][]{Quinn1984, Hernquist&Quinn1988}. 
Accreted stars have a spread in energies and a corresponding spread in orbital periods.
Once they are stripped, stars with shorter periods will begin to lead those with longer periods, creating a wrap in phase space \citep[see][]{Johnston1998, Amorisco2015}. Stars at the apocenter have relative radial velocities ($v_r$) close to zero, so that
stellar shells are visible in the first two-columns as caustics in phase space. 
In-situ material, on the other hand, is situated closer to the center of the host galaxy and well-mixed.

As already suggested by \cite{Quinn1984}, the phase space representation of shell galaxies also allows us to probe the age of the shells. Young systems in which accretion of the satellite happened recently only exhibit one or two wraps, with the most bound particles in the satellite still concentrated in both configuration and phase space. In time, phase wraps become interleaved in radius and more shells are seen on alternating sides of the galaxy in configuration space. Therefore, the number of wraps provides an estimate of the merger lookback time.
At later times, the shells become more diffused, as stars suffer phase mixing in the gravitational potential of the new host.

\begin{figure*}
\vspace{-0em}\includegraphics[width=\textwidth]{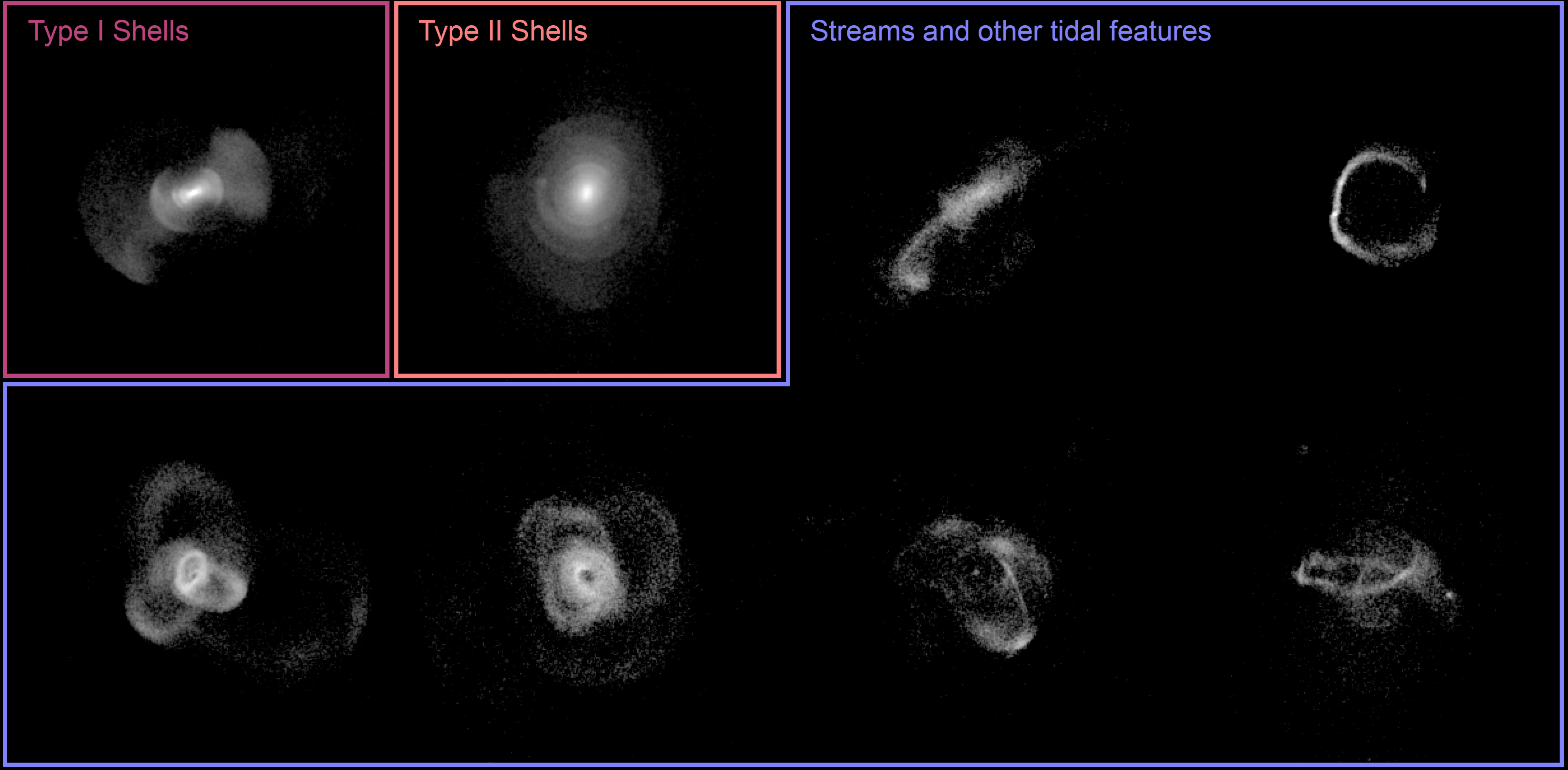}
\caption{Examples of different tidal features in Illustris, with each panel showing the stellar surface mass density of $z=0$ stars that have been stripped from the same progenitor galaxy. The two galaxies in the upper left corner correspond to shell structures and they are the focus of this paper. Type I shells are aligned with the major axis of the galaxy, while type II shells have wider opening angles and relatively random orientations in space. Alongside shells, we present examples of other types of tidal features that are the result of satellites accreted on orbits with higher angular momentum. In particular, the features in the second row correspond to wrappings. 
Differently from shells, very few of the stripped stars that compose the wrappings are found near the center of the host galaxy. Moreover, the resulting wraps are thin in the direction perpendicular to the orbital plane. In this paper, we consider only shell features, without distinguishing among different shell types (I, II, or III).}
\label{fig:nonshells}
\end{figure*}

In Figure~\ref{fig:nonshells}, we compare shells to examples of other types of tidal features in our simulation.
In this series of images, we isolate the $z=0$ distribution of stars that were stripped from the same common progenitor galaxy.
Type I shells are easily recognizable due to the alignment of the shells with a common axis, whereas Type II shells are randomly distributed in space and cover wider opening angles.
We do not require here a systematic classification of the continuous morphologies of other types of tidal features. We just note that angular momentum drives the formation of qualitatively different morphologies. While shells form through near-radial mergers, satellites accreted on orbits with higher angular momentum will preferentially form tidal streams.
As shown in Figure~\ref{fig:nonshells}, streams can have a range of morphologies. In particular, satellites accreted on orbits with intermediate angular momentum will form stellar `wrappings' like those showcased in the bottom row of Figure~\ref{fig:nonshells}. Observed from certain viewing angles, these features can resemble shells, however, they remain stream-like in that they are thin in the direction perpendicular to the orbital plane. Additionally, very few of the stars in these `wraps' are found at the center of the host galaxy.

Based on the arguments above, we proceed to define stellar shells in our simulation as low surface brightness tidal debris that form interleaved caustics, most often observed on both sides of the galaxy center. Our sample includes Type I, II, and III shells, and we do not distinguish among these types in this work. In configuration space, we identify shells based on sharp arc-like features at large galactocentric distances, while in phase space, we identify shells as caustics peaking close to $v_r \simeq 0$. 

\subsubsection{Step 1: Visual Identification}
\label{subsubsec:step1Visual}

\begin{figure*}
\centering
\includegraphics[width=.96\textwidth]{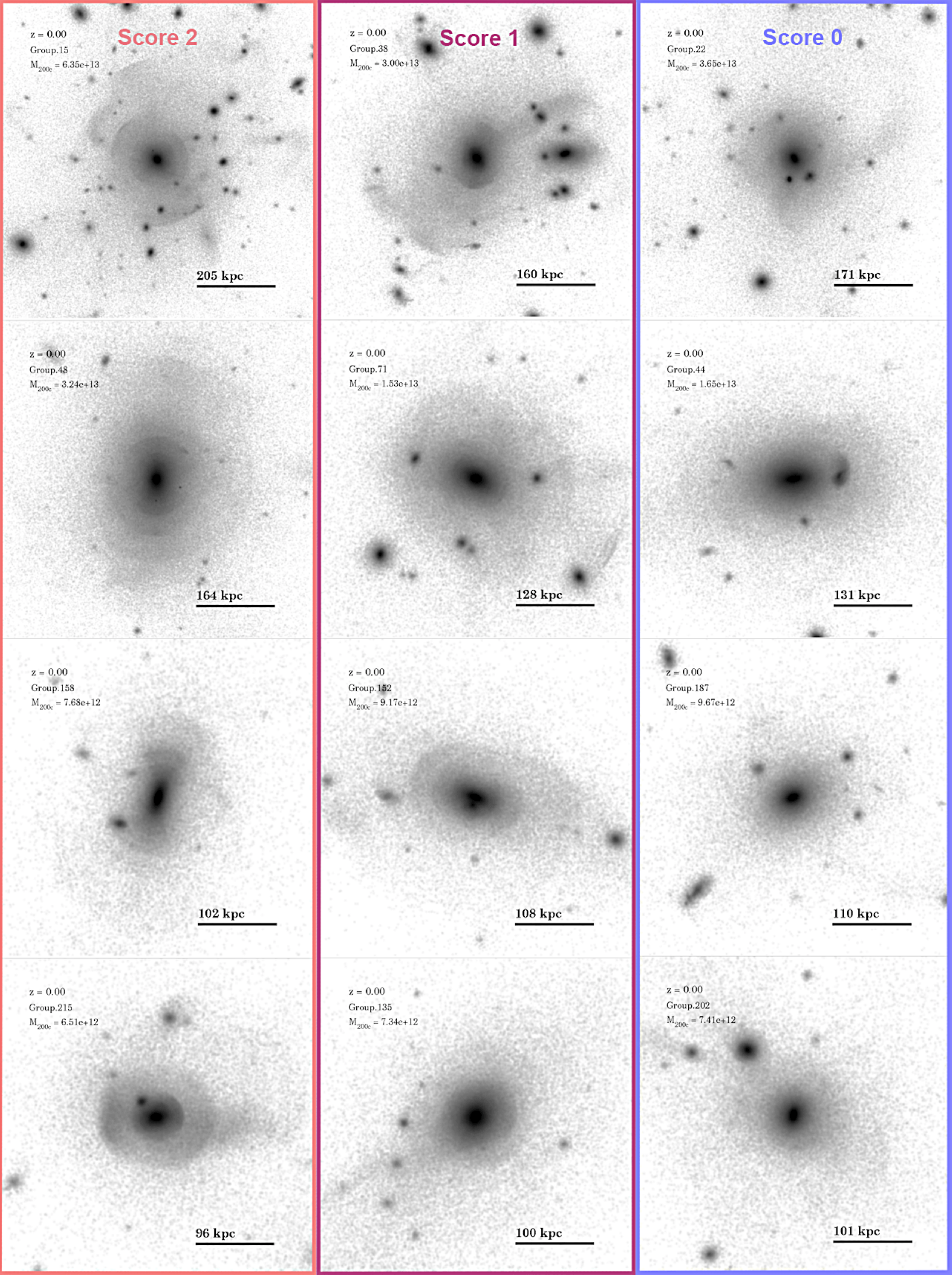}
\caption{In selecting galaxies with stellar shells, we employ a scoring system between $2$ and $0$. The images above are examples of galaxies with similar $z=0$ mass. From left to right, each galaxy received a score of 2 (exhibiting two or more well-defined shells), a score of 1 (for those cases when we detect 1-2 shell-like structures), or a score of 0 (no shell detection). The mass quoted in each figure is $\mathrm{M}_{200\mathrm{crit}}$ in solar mass units. }
\label{fig:scoring}
\end{figure*}

\begin{figure*}
\vspace{-0em}\includegraphics[width=\textwidth]{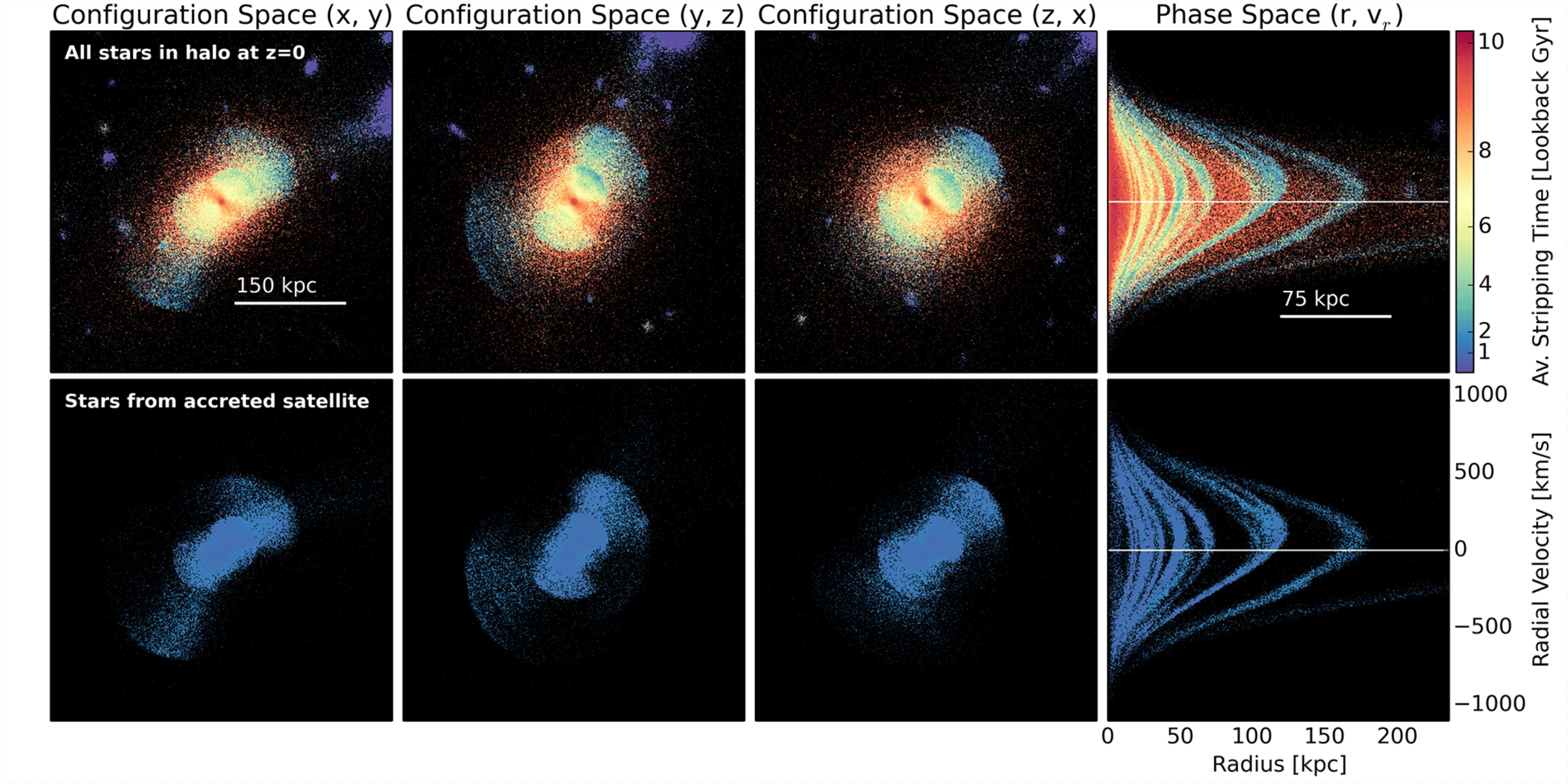}
\caption{Stellar surface density maps in configuration space (first three columns) and phase space (last column) for an entire halo with stellar shells at redshift $z=0$ (top row). The stellar particles are colored based on their stripping time, with red stars stripped a very long time ago, while purple stars correspond to satellites that have yet to be stripped by $z=0$. The bottom row shows $z=0$ stars that have been accreted from the same common progenitor. These stars are responsible for the shell structure observed in the top panels and they have a fairly recent stripping time (within the past $\sim$2 Gyr). Note that the color of each pixel corresponds to the average stripping time of all stars therein; as a result, the shells in the top row appear lighter in color than those in the bottom row due to other stars in the same pixel that have earlier stripping times.}
\label{fig:strippingTime}
\end{figure*}

We identify galaxies with shells in Illustris using a two-step approach: 1) we use stellar surface density maps to visually identify shells and 2) we find the progenitor galaxies forming shells using stellar history catalogs.

For the visual identification step, we put ourselves in the best possible position for detecting all shells in the Illustris simulation box. We use stellar surface density maps such as those in Figure~\ref{fig:obsvsIllustris} and we do not apply a surface brightness cut. We look at each galaxy in our sample using three projections ($x-y$, $y-z$, $z-x$), since some shell systems are visible in only one or two projections, depending on the trajectory of the accreted satellite (see discussion in Appendix~\ref{sec:projectionEffects}). We also use two contrast levels: high central contrast helps us identify outer shells, while lower contrast is better for identifying shells at small galactocentric distances.

Similar to previous observational surveys that classify galaxies based on low surface brightness tidal features such as shells, streams, plumes, tails, etc. \citep[e.g.,][]{Atkinsonetal2013, Ducetal2014}, we use visual identification and employ a scoring system allowing us to differentiate between clear shell detections and more faint shell-like structures.
Three members of our team give scores to each galaxy in our sample on a scale from $0-2$. We give a score of $2$ for galaxies that show two or more well-defined shells, a score of $1$ for intermediate galaxies that exhibit one or two shell-like structures, and a score of 0 for those galaxies that do not exhibit shells. Examples of galaxies that received each of these scores are presented in Figure~\ref{fig:scoring}.
The galaxies' masses decrease from top to bottom in the figure, with galaxies in the same row having roughly similar masses.
We graded six different image stamps per galaxy (three different projections, two contrast levels).
The score distribution averaged over all six stamps and three sets of grades is included in Figure~\ref{fig:scoreDist}.

\subsubsection{Step 2:  Identifying Progenitors Forming Shells}
\label{subsubsec:step2Confirm}

Based on the stellar history catalogs described in \S\ref{subsec:stellarHistoryCatalogs}, we then develop a set of postprocessing tools that allow us to efficiently track the parent satellite of each star. 
This is similar to building a merger tree catalog, except that we are only interested in the 
direct progenitors of the host galaxies, in order to assign 
stars inside the shells to their corresponding progenitors. 
\cite{Rodriguez-Gomezetal2016} previously examined the fraction of in-situ and ex-situ material in Illustris galaxies, using merger trees to investigate the origin of individual star particles in the simulation. 
The scope of studying the dynamics of shell formation requires us to go into more depth, with the stellar history catalogs including detailed information about the stars' dynamics and parent satellite at several key moments (formation, accretion, stripping).
This is a computationally demanding task, and as a result, we only save information for star particles that are inside the $z=0$ stellar halos of galaxies identified through step 1 as potential shell galaxies. 
We reconstruct the detailed assembly histories of shell galaxies using the additional postprocessing tools described below. 

For each snapshot in the simulation, we order all satellites that provide stars to the final host based on the number of star particles that were accreted from that satellite at the given time. 
We investigate progenitors with at least $10^4$ star particles, corresponding to a stellar mass $\mathrm{M}_{\mathrm{stars,tot }} \gtrsim 9 \times 10^9 \, \mathrm{M}_\odot$. As discussed in Appendix~\ref{sec:resolution}, we reckon this does not limit our ability to identify shells.
Since some of the ex-situ stars might form after their parent satellite has already crossed inside the virial radius ($\mathrm{R}_{\mathrm{vir}}$) of the final host, the accretion of a satellite can extend over a series of snapshots. We take this effect into account
by linking the accretion time of each progenitor
to the snapshot when most of the stars in the satellite are being accreted by the host.     
Thus, we obtain a systematic sample of all the satellites accreted by the final host and the detailed histories of each star inside these progenitors. 

Then, we investigate stars that belonged to a common satellite prior to being accreted by the host, and we trace the position of these stars in configuration and phase space at $z=0$.
As demonstrated in Figure~\ref{fig:strippingTime}, this approach successfully recovers the shells observed at $z=0$, and allows us to identify the satellites that contribute stars to the shell structures.
In Figure~\ref{fig:strippingTime}, we 
color stars based on their stripping time, with red stars being stripped a long time ago, and stars colored in blue being stripped more recently.
Purple indicates stars that have yet to be stripped, i.e. stars belonging to satellites of the central galaxy.
The bottom panel of Figure~\ref{fig:strippingTime} shows only stars that were stripped from the same progenitor. By comparing the shells in the two rows, we identify this merger event as the source of the shells observed at $z=0$. After checking if any of the other progenitors of this galaxy have stars arranged in shells, we confirm all the merger events responsible for the $z=0$ shells.

We use this procedure to confirm the nature of the candidate shells identified in step 1 (\S\ref{subsubsec:step1Visual}).
In the rest of the paper, we refer to those galaxies first marked through the visual identification step and then confirmed through the second step as "shell galaxies". 
We find that 39 of the 220 galaxies in our initial sample exhibit shells at $z=0$.

\section{The Distribution of Shell Galaxies at z=0}
\label{sec:distShells}

\subsection{The Mass Distribution and Fraction of Shell Galaxies}
\label{subsec:fraction}

\begin{figure*}
\vspace{-0em}\includegraphics[width=\textwidth]{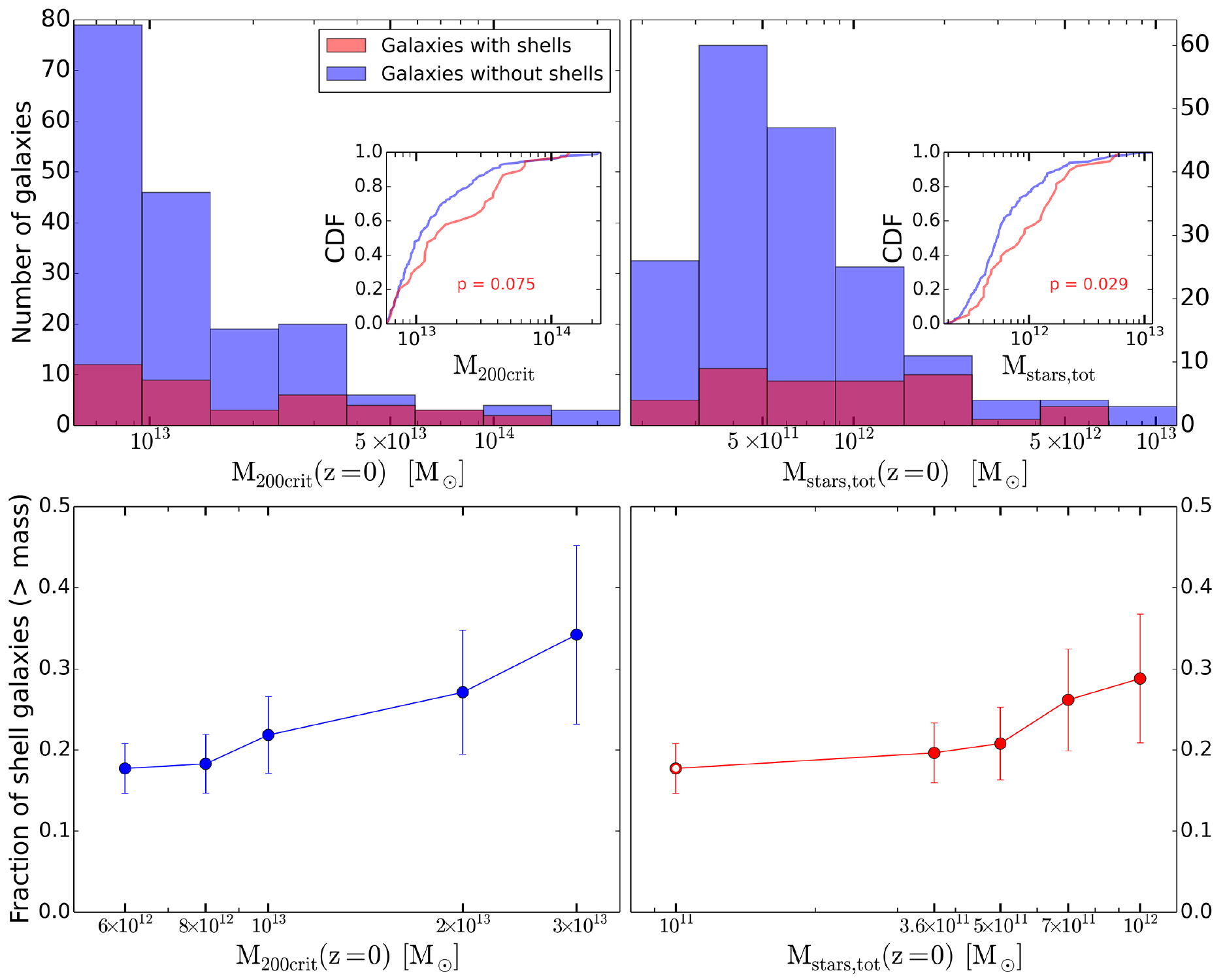}
\caption{\textit{Top row}: Mass distribution of shell galaxies (in red) and galaxies without shells (in blue). The left panel corresponds to $\mathrm{M}_{200\mathrm{crit}}$, while the right panel shows the total stellar mass of the entire halo. The insets in both panels show cumulative distribution functions and the p-value from two-sample KS tests, quantifying the likelihood that galaxies with and without shells come from the same underlying distribution. 
Galaxies with higher masses are more likely to display stellar shells.
\textit{Bottom row}: Fraction of galaxies exhibiting shells at redshift z=0, when we restrict our sample above a given mass in $\mathrm{M}_{200\mathrm{crit}}$ (blue points on the left) or above a given stellar mass $\mathrm{M}_{\mathrm{stars}}$ (red points on the right). Our complete sample of galaxies includes all 220 galaxies in the Illustris simulation with a mass $\mathrm{M}_{200\mathrm{crit}} >6 \times 10^{12}\, \mathrm{M}_\odot$, and we find that 39 of those galaxies have shells (i.e. about $18\% \pm 3\%$ of all galaxies in our sample). The fraction of shell galaxies increases monotonically with increasing total/stellar mass. Our galaxy sample is complete for stellar masses $\mathrm{M}_{\mathrm{stars}} > 3.6 \times 10^{11}\, \mathrm{M}_\odot$, but we continue to see a similar trend in the fraction of shell galaxies vs. $\mathrm{M}_{\mathrm{stars}}$ down to the minimum stellar masses in our sample ($\mathrm{M}_{\mathrm{stars}} \gtrsim 10^{11}\, \mathrm{M}_\odot$ - marked by an empty circle).}
\label{fig:massDistofShells}
\end{figure*}

In this subsection we investigate the mass distribution of shell galaxies in both the virial mass of the host, $\mathrm{M}_{\mathrm{200crit}}$, and total stellar mass of the halo, $\mathrm{M}_{\mathrm{stars,tot}}$.
The top panels of Figure \ref{fig:massDistofShells} include histograms for the mass distribution of $z=0$ shell galaxies, accompanied by cumulative distribution functions in the insets of each panel.
Shell galaxies are shown in red, while galaxies without shells are depicted in blue. We find that shell galaxies tend to have a more flat mass distribution, indicating that higher mass galaxies have a somewhat higher likelihood to form shells visible at $z=0$.
We explore this using two-sample Kolmogorov-Smirnov tests and find that the probability that shell galaxies and galaxies without shells have the same parent mass distribution is rather low: p-values of $p = 0.029$ and $p = 0.075$ for the stellar mass and total mass distributions, respectively.
While larger sample sizes would be desirable, these results suggest that higher-mass galaxies display $z=0$ shells more frequently than their less massive counterparts.
Discussions on the physics potentially responsible for this trend are included in Section~\ref{sec:discussion}.

The two steps described in the previous section select 39 galaxies with stellar shells out of an initial sample of 220 massive galaxies in Illustris ($\mathrm{M}_{\mathrm{200crit}} > 6 \times 10^{12}\, \mathrm{M}_{\odot}, \; \mathrm{M}_{\mathrm{stars}} > 10^{11}\, \mathrm{M}_{\odot}$). 
This corresponds to $18\% \pm 3\%$ shell galaxies in our total sample of galaxies. 
The bottom panels in Figure~\ref{fig:massDistofShells} show how the fraction of shell galaxies depends on the total (stellar) mass cut applied to the galaxies in our sample (here, stellar mass is computed as the total mass of all stars in the halo). 
The left panel shows that the fraction of shells increases with increasing mass cuts in $\mathrm{M}_{200\mathrm{crit}}$, varying from $18\% \pm 3\%$ for $\mathrm{M}_{200\mathrm{crit}} > 6 \times 10^{12}\, \mathrm{M}_\odot$ up to $34 \% \pm 11\%$ for $\mathrm{M}_{200\mathrm{crit}} > 3 \times 10^{13}\, \mathrm{M}_\odot$. 
Since stellar and total masses are correlated, we see a similar trend in the right panel, with the fraction of shells increasing with mass up to $29\% \pm 8\%$ for $\mathrm{M}_{\mathrm{stars}} > 10^{12}\, \mathrm{M}_\odot$. 
Galaxies were selected based on a mass cut in $\mathrm{M}_{200\mathrm{crit}}$ and as a result, our sample is only complete above $\mathrm{M}_{\mathrm{stars,tot}} > 3.6 \times 10^{11}\, \mathrm{M}_\odot$. Nonetheless, the trend of decreasing fraction of shell galaxies with decreasing stellar mass cut continues down to a mass cut of $\mathrm{M}_{\mathrm{stars,tot}} \gtrsim 10^{11}\, \mathrm{M}_\odot$ (marked by an empty red circle), which corresponds to the lowest stellar mass in our sample.
Errors bars were computed using error propagation for the fraction of shells defined as $f = \mathrm{N}_{\mathrm{shells}}/\mathrm{N}_{\mathrm{galaxies}}$, considering Poisson errors for both $\delta \mathrm{N}_{\mathrm{shells}} = \sqrt{\mathrm{N}_{\mathrm{shells}}}$ and $\delta\mathrm{N}_{\mathrm{galaxies}} = \sqrt{\mathrm{N}_{\mathrm{galaxies}}}$.

\subsection{Redshift Evolution of the Fraction of Shell Galaxies}
\label{subsec:redshift}

\begin{figure}
\vspace{-0em}\includegraphics[width=\columnwidth]{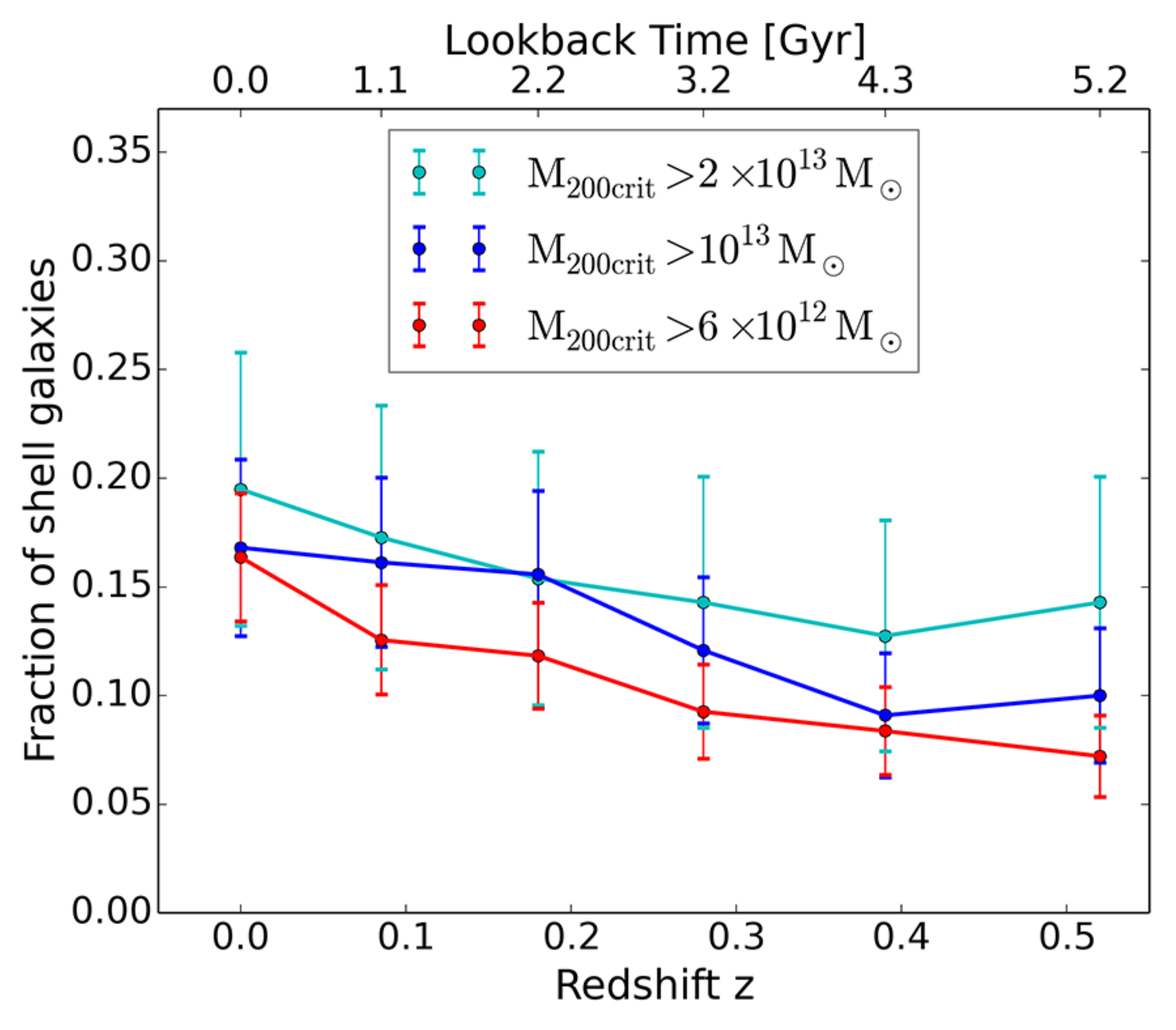}
\caption{Redshift evolution of the fraction of shell galaxies with $\mathrm{M}_{\mathrm{200crit}}(z)$ $> 2 \times 10^{13}\, \mathrm{M}_{\odot}$ (cyan), $ > 10^{13}\, \mathrm{M}_{\odot}$ (blue),  $> 6 \times 10^{12}\, \mathrm{M}_{\odot}$ (red), with $\mathrm{M}_{\mathrm{200crit}}$ measured at each given redshift z. Thus, the mass selections for each line are cumulative, and they correspond to galaxies above the given mass cut at that redshift. 
Across all three different mass cuts, the fraction of shell galaxies decreases with redshift, with galaxies $\mathrm{M}_{\mathrm{200crit}}> 6 \times 10^{12}\, \mathrm{M}_{\odot}$ being roughly two times less likely to have shells at $z=0.5$ versus $z=0$. }
\label{fig:redshiftEvolution}
\end{figure}

In order to test how the fraction of shell galaxies evolves with redshift, we apply the first identification step described in \S \ref{subsec:galaxySample} to galaxies above $\mathrm{M}_{\mathrm{200crit}} > 6 \times 10^{12}\, \mathrm{M}_{\odot}$ at six different redshifts roughly equally spread out in lookback time: $z = 0$, $0.085$, $0.18$, $0.28$, $0.39$, $0.52$. 
Due to the high computational expense, we do not produce stellar history catalogs (second identification step) to confirm our shells at higher redshifts. Instead, we use our sample of $z=0$ shell galaxies to
estimate the score above which shells are accurately identified after the first step (see Figure \ref{fig:scoreDist} for a quantitative estimate). 

The resulting redshift evolution of the fraction of shells is shown in Figure \ref{fig:redshiftEvolution} for three different mass cuts $\mathrm{M}_{\mathrm{200crit}}(z) > 2 \times 10^{13}\, \mathrm{M}_{\odot}$, $ > 10^{13}\, \mathrm{M}_{\odot}$, $> 6 \times 10^{12} \, \mathrm{M}_{\odot}$, applied at each redshift z. 
Error bars are computed using error propagation: $\frac{\delta f}{f} = \sqrt{ \left(\frac{\delta \mathrm{N}_{\mathrm{shells}}}{\mathrm{N}_{\mathrm{shells}}} \right)^2 + \left(\frac{\delta \mathrm{N}_{\mathrm{total}}}{\mathrm{N}_{\mathrm{total}}} \right)^2}$, leading to $\delta f = \sqrt{\frac{\mathrm{N}_{\mathrm{shells}}}{\mathrm{N}_{\mathrm{total}}^2} \left( 1 +\frac{\mathrm{N}_{\mathrm{shells}}}{\mathrm{N}_{\mathrm{total}}} \right)  }$. 
For our sample of massive ellipticals, the fraction of shells decreases monotonically with redshift, getting as low as $7\% \pm 2\%$ for galaxies with $\mathrm{M}_{\mathrm{200crit}}(z=0.5) > 6 \times 10^{12}\, \mathrm{M}_{\odot}$. 
The overall trend of decreasing fraction of shells at higher redshifts is consistent across the different mass cuts studied in Figure~\ref{fig:redshiftEvolution}, although it could be different for less massive galaxies.
In Section~\ref{sec:discussion}, we discuss how depositing the shell material closer to the center of galaxies, together with shorter phase-mixing times, could make it more difficult to detect shell galaxies at higher redshifts.

\section{The Merger Events That Produce Shells}
\label{sec:progenitors}

In this section, we compare those progenitors that successfully form $z=0$ shells to all the other satellites accreted by the main host galaxy. 
Our goal is to find the order-zero recipe, involving the smallest number of parameters, that describes the merger events responsible for the shells at redshift $z=0$.

\subsection{Recipe for Forming Shells in High-Mass Galaxies}
\label{subsec:recipe}

\begin{figure*}
\vspace{-0em}\includegraphics[width=1.02\textwidth]{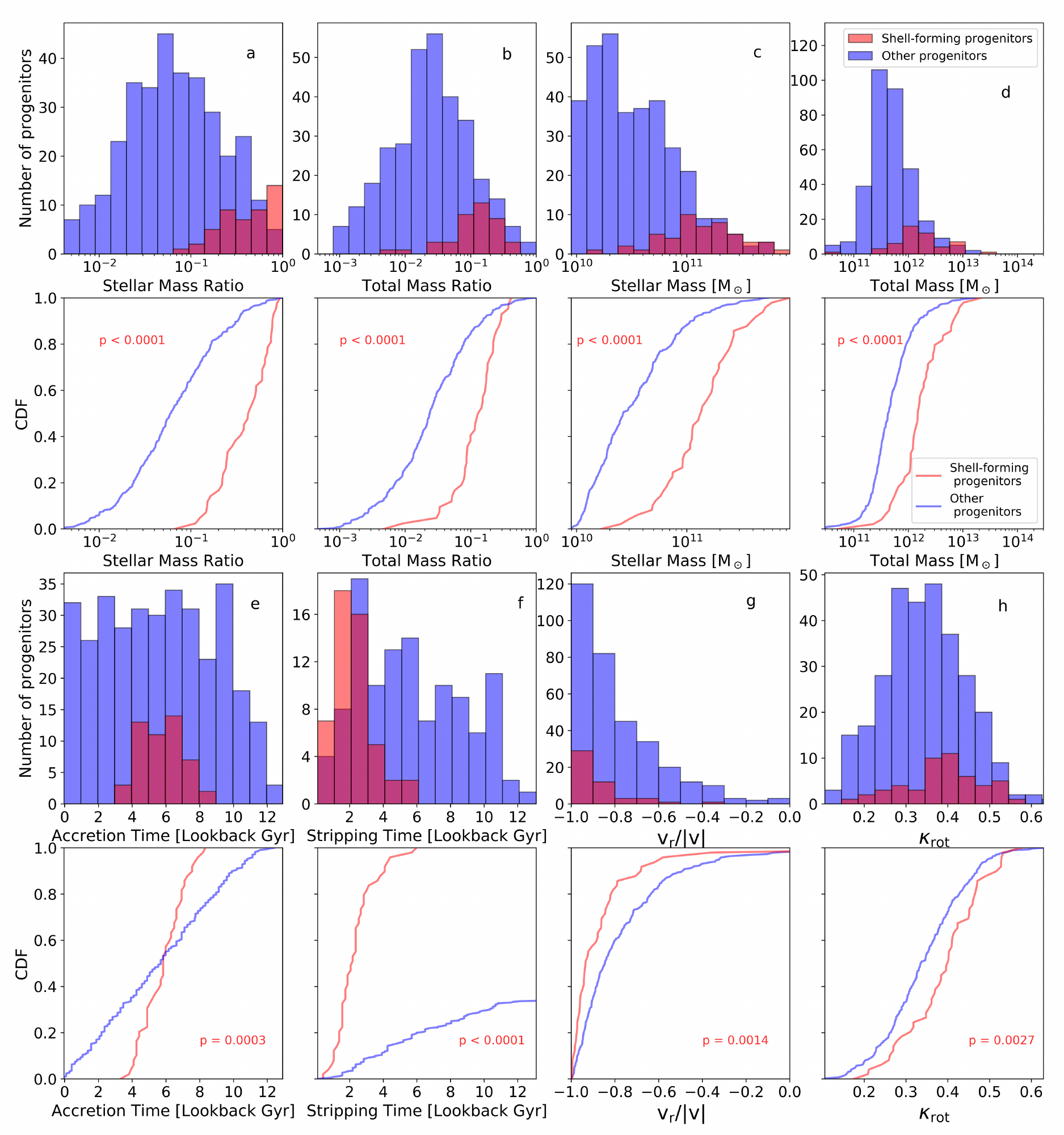}
\caption{Analysis of how shell-forming progenitors differ with respect to the overall satellite population. We study the stellar and total mass ratio (panels a and b), stellar and total mass (c and d), accretion and stripping time (e and f), radial velocity (g) and circularity (h) of each progenitor. The first and third row show histograms, while the second and fourth row show the corresponding cumulative distribution functions for each property. The insets represent the KS-test p-value between the shell-forming progenitors and overall progenitor population. Shell-forming progenitors have higher stellar and total mass ratios than most progenitors, they were accreted at intermediate times ($\sim$4 -- 8 Gyr ago) and stripped within the last $\sim$6 Gyr. Shell-forming progenitors come on more radial orbits (i.e. $\mathrm{v}_\mathrm{r}/|\mathrm{v}| \simeq -1$) and on average, they have higher $\kappa_{\mathrm{rot}}$ than the overall progenitors. Note that the CDF for stripping time of the overall progenitors doesn't reach one since some of the progenitors have yet to be stripped by $z=0$. }
\label{fig:progenitors}
\end{figure*}

For each one of our 39 shell galaxies, we identify all the satellites\footnote{We only consider satellites with more than $10000$ star particles, as discussed in Appendix~\ref{sec:resolution}.} with $\mathrm{M}_{\mathrm{stars,tot}} \gtrsim 9 \times 10^9\, \mathrm{M}_\odot$ accreted by $z=0$ and classify these satellites based on whether they have successfully formed shells or not. 
In Figure~\ref{fig:progenitors} we investigate whether shell-forming progenitors (shown in red) have significantly different properties from those satellites that do not form shells (shown in blue). Alongside histograms, Figure~\ref{fig:progenitors} includes cumulative distribution functions (CDFs) for each physical quantity investigated, and results from two-sample Kolmogorov-Smirnov (KS) tests, quantifying the likelihood that the sample of shell-forming progenitors and the sample of all other progenitors have been drawn from the same underlying probability distribution. 

Panels (\textit{a}) and (\textit{b}) of Figure~\ref{fig:progenitors} show the distributions of stellar and total mass ratios ($\mu_{\mathrm{stars}} = \mathrm{M}^{\mathrm{stars}}_{\mathrm{sat}}/\mathrm{M}^{\mathrm{stars}}_{\mathrm{host}}$ and $\mu_{\mathrm{total}} = \mathrm{M}^{\mathrm{total}}_{\mathrm{sat}}/\mathrm{M}^{\mathrm{total}}_{\mathrm{host}}$), where $\mathrm{M}^{\mathrm{stars}}$ and $\mathrm{M}^{\mathrm{total}}$ correspond to the mass of all stellar particles/all particles gravitationally bound to the respective halo. These mass ratios are computed at accretion time ($t_{\mathrm{acc}}$), i.e. when the progenitor comes within one virial radius of the host\footnote{Note that we exclude from the stellar and total mass ratio panels those very few satellites for which our mass ratios measured at $t_{\mathrm{acc}}$ are \mbox{$\mu>1$}. 
This happens in the case of major mergers, when the main host is about to undergo mergers with massive satellites already inside its virial radius or when the 
SUBFIND algorithm has issues assigning stellar particles to their respective subhalos. Removing these few occurrences from our sample has a negligible effect on the p-value of the K-S tests.}.
In this paper, we refer to approximately major mergers for which \mbox{$\mu = \mathrm{M}_{\mathrm{sat}}/\mathrm{M}_{\mathrm{host}} \gtrsim 1/10$} as \textit{high} mass ratio mergers.
Correspondingly, we will use the terms \textit{low} mass ratio mergers and minor mergers interchangeably.
We find that shell-forming progenitors are involved in relatively major mergers, with almost all corresponding merger events having a stellar mass ratio  $\mu_{\mathrm{stars}} \gtrsim 1/10$. 
In addition, more than $50\%$ of the shell-forming progenitors have $\mu_{\mathrm{stars}} > 1/3$. 
While the mean total mass ratio ($\mu_{\mathrm{total}}$) for shell-forming progenitors is lower than the mean $\mu_{\mathrm{stars}}$,
both the $\mu_{\mathrm{total}}$ and $\mu_{\mathrm{stars}}$ distributions are 
significantly biased towards more major mergers. 
As seen in panels (\textit{c}) and (\textit{d}) of Figure~\ref{fig:progenitors}: the progenitors' stellar and total masses at $t_{\mathrm{acc}}$ are biased high,
with about half of all the progenitors above $M_{\mathrm{stars}} = 10^{11}\, \mathrm{M}_\odot$ forming $z=0$ shells.

The time when the satellites are being accreted and then stripped by the main host is also an essential ingredient in determining whether they form detectable shells at $z=0$. In panels (\textit{e}) and (\textit{f}) of Figure~\ref{fig:progenitors}, we find that shell-forming progenitors are accreted on average between 4.2 and 7.6 Gyr ago ($10\%$ and $90\%$ percentiles, respectively), and they were stripped between $\sim$1 and 4 Gyr ago. 
Satellites accreted too late ($t_{\mathrm{acc}} \lesssim 4$ lookback Gyr) do not have enough time to be stripped and form shells by $z=0$.
Some of the satellites accreted too early ($t_{\mathrm{acc}} \gtrsim 8$ lookback Gyr) could have produced shells, but those features most likely phase-mixed by $z=0$. 
The most likely stripping time\footnote{The host galaxy is stripping stars from the accreted progenitor over a few hundred Myr, so there are several possible definitions for the "stripping time" ($t_{\mathrm{strip}}$) of a satellite. We choose to define the stripping time as the median $t_{\mathrm{strip}}$ for all the stars in the satellite, which is in almost all cases identical to choosing the simulation snapshot when most stars are being stripped. } for stars in shells is about 2.1 Gyr ago.
For those stripping events that peaked at $t_{\mathrm{strip}} \gtrsim 4$ lookback Gyr, the shells have already started to phase-mix, making them increasingly more difficult to detect. A more detailed discussion of these effects is included in Section~\ref{sec:discussion}.

Shell-forming progenitors also have more radial infall orbits. In panel (\textit{g}) of Figure~\ref{fig:progenitors}, we measure the radial velocity ratio ($v_r/|v|$) at $t_{\mathrm{acc}}$, defined as the radial component of the relative velocity of the satellite with respect to the host, normalized to the modulus of the relative velocity. Satellites that form shells have a median radial velocity ratio $v_r/|v| = -0.94$. Moreover, almost $90\%$ of the shell-forming satellites come on orbits where the radial velocity component dominates over the tangential one (i.e., $v_r/|v|  < - 1/\sqrt{2} = - 0.71$). We measure this radial velocity ratio at the accretion time and we expect the orbits of stripped satellites to become increasingly more radial during the final infall due to dynamical friction (see \S \ref{subsec:dicussProgenitors}).

In order to quantify the morphology of the progenitors (i.e., distinguish between discs and spheroids), we compute the $\kappa_{\mathrm{rot}}$ parameter as introduced by \cite{Salesetal2012}:
\begin{eqnarray}
\kappa_\mathrm{rot} = \frac{K_{\mathrm{rot}}}{K} = \frac{1}{K} \sum_i \frac{1}{2} m_i \left( \frac{j_{z,i}}{R_i} \right)^2,
\end{eqnarray} \label{eqn:kappa}
i.e., the fraction of kinetic energy invested in ordered rotation. Here, $K$ is the total kinetic energy of all stars in the galaxy, while the summation is done over each star's mass $m_i$, specific angular momentum along the z-axis $j_{z,i}$ (oriented along the direction of the total angular momentum), and projected radius $R_i$. As discussed in \cite{Salesetal2012}, $\kappa_{\mathrm{rot}}$ is a good proxy for the amount of rotational support in a galaxy: we expect $\kappa_{\mathrm{rot}} = 1/3$ for an isotropic stellar system where all three velocity components are equal at any radius, while $\kappa_{\mathrm{rot}}$ would be 1 for a perfectly cold rotating disk.
Real disks do not reach values as high as 1, but $\kappa_{\mathrm{rot}}$ correlates well with the fraction of stars with stellar orbital parameter $\varepsilon_j = j_z / j_{\mathrm{circ}}(E) > 0.5$ \citep[$\varepsilon_j$ is the ratio between the component of the angular momentum aligned with the total angular momentum of the galaxy and the angular momentum of a circular orbit with binding energy E; see, e.g.,][]{Abadietal2003, Marinaccietal2014}. 
As shown in panel (\textit{h}), shell-forming progenitors tend to have a higher mean $\kappa_{\mathrm{rot}}$ (i.e., they are more rotation-dominated) than the overall population of accreted satellites, but there is large scatter in both distributions. This result suggests that colder material might be more likely to form shells, as hinted by earlier studies such as \cite{Quinn1984} and \cite{Hernquist&Spergel1992}.

\begin{figure*}
\vspace{-0em}\includegraphics[width=0.8\textwidth]{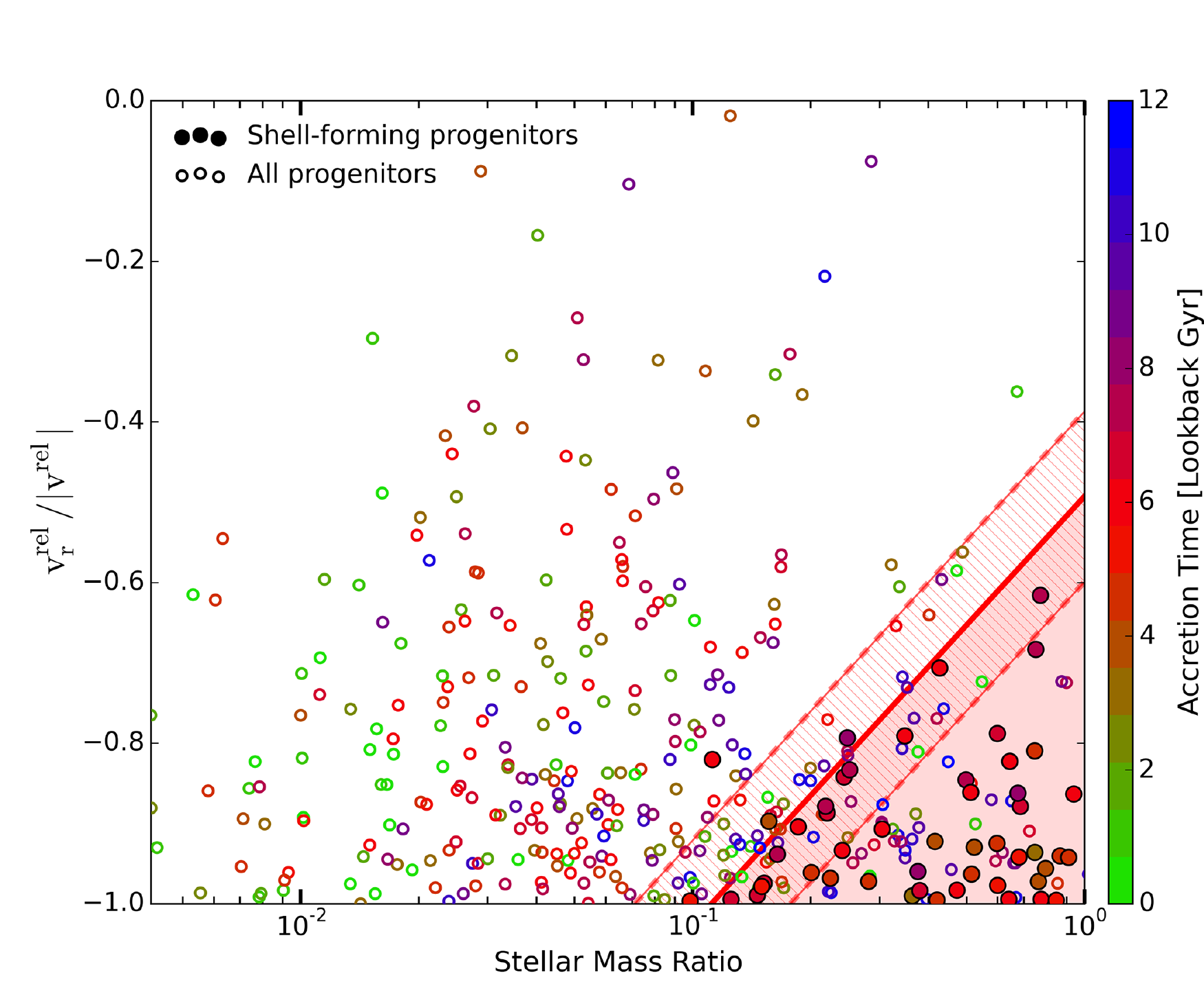}
\caption{Comparison between shell-forming progenitors (marked with full circles) and all the other progenitors (empty circles), in the three-dimensional parameter space radial velocity - stellar mass ratio - accretion time. The accretion time of each satellite is indicated by the color of its corresponding circle: satellites accreted in the very recent past are colored in green, while satellites accreted earlier are colored in blue. We find that progenitors responsible for $z=0$ shells were accreted $\sim$4 -- 8 Gyr ago (colored in red). They have high stellar mass ratios ($\mu_{\mathrm{stars}} = \mathrm{M}_{\mathrm{sat}}^{\mathrm{stars}} / \mathrm{M}_{\mathrm{host}}^{\mathrm{stars}} > 0.1 $) and roughly radial orbits. The preferred region of parameter space for shell-forming progenitors is marked by the red triangle in the lower-right corner of the plot. As shown in this figure, the triplet (high $\mu_\mathrm{stars}$, intermediate $t_{\mathrm{acc}}$, low $v_r/|v|$) provides a zero-order recipe for characterizing the satellites that successfully form $z=0$ shells in high-mass hosts. }
\label{fig:threeDim}
\end{figure*}

From the eight parameters studied in Figure~\ref{fig:progenitors}, we pick three of the parameters with lowest p-values\footnote{Stripping time has a lower p-value than accretion time, but this is driven by the fact that many of the satellites not forming shells have not been stripped by $z=0$. Similarly, we choose to only include stellar mass ratio, as stellar and total masses of the satellites are correlated.}: stellar mass ratio, accretion time, and radial velocity ratio. We argue that these three parameters alone can provide a simple recipe for most satellites forming shells in high-mass 
galaxies at $z=0$. Figure~\ref{fig:threeDim} shows the distribution of shell-forming satellites (full circles) compared to all the other progenitors (open circles). We find that shell-forming progenitors correspond to satellites accreted with high stellar mass ratios ($\mu_{\mathrm{stars}} \gtrsim 0.1$) on approximately radial orbits, about $4-8$ Gyr ago (red-colored circles). In the current study, we only consider satellites with more than $10^4$ star particles, allowing us to resolve well-sampled stellar shells. 
This limit does \textit{not} affect our quantitative findings: as demonstrated in Appendix~\ref{sec:resolution}, Figure~\ref{fig:resolution}, our cut at $10^4$ star particles per satellite would have still allowed us to find smaller shell-forming progenitor mass ratios than the ones we found for Illustris shell galaxies. 
However, it so occurs that progenitors with smaller mass ratios (that are just above or at our star particle number cut) do not form shells.

We use balanced logistic regression \citep[see][]{Cox1958} in order to disentangle the correlation between the radial velocity ratio and the stellar mass ratio of the satellites. Our methods are described in more detail in Appendix~\ref{sec:logisticRegression}. The red line in Figure~\ref{fig:threeDim} separates regions of the parameter space ($v_r, \mu_{\mathrm{stars}}$) where we are equally likely to find shell-forming and non-shell-forming satellites, and \mbox{1 $\sigma$} bounds are marked with dashed lines. 
The results presented in Figure~\ref{fig:threeDim} show that for satellites with higher mass ratios (i.e., closer to 1:1 mergers), the radial velocity condition becomes less stringent:
while relatively massive satellites can still form shells when they are accreted with $v_r/|v| \lesssim -0.5$, less massive satellites require more finely tuned infall orbits, with very low angular momentum.
This effect is due to the increasing efficiency of dynamical friction in radializing the orbits of major mergers and it is further discussed in \S\ref{subsec:dicussProgenitors}. 

In special cases, 
shell-forming progenitors can either lie outside the preferred red triangle in Figure~\ref{fig:threeDim} or vice versa,
satellites inside this preferred region in ($v_r/|v|$, $\mu_{\mathrm{stars}}$) might fail to form shells. 
For example, the full circle positioned above the $1\sigma$ bounds from the preferred region in Figure~\ref{fig:threeDim} 
corresponds to a shell-forming satellite accreted while the central host galaxy is undergoing a major merger.
Measuring its
radial velocity relative to the center of mass of the halo reduces the radial velocity ratio from $v_r/|v| = -0.82$ to $v_r/|v| = -0.90$, and thus brings this satellite within the $1\sigma$ bounds. Another scenario when a progenitor can lie outside the triangle corresponds to satellites accompanied by their own smaller satellites (see discussion in \S\ref{subsec:satofsat} about satellites-of-satellites).

We also identify several satellites in Figure~\ref{fig:threeDim} that lie in the preferred region of the three-dimensional parameter space ($v_r/|v|$, $\mu_{\mathrm{stars}}$,  $t_{\mathrm{acc}}$) but they fail to produce shells visible at redshift $z=0$.
By investigating each of these satellites, we find two different explanations:

1) Some of these progenitors successfully form shells, as expected due to their high stellar mass ratios and low angular momentum orbits ($v_r/|v|$ close to $-1$ at $t_{\mathrm{acc}}$). However, the resulting shells phase mix before $z=0$. This can happen because either:
(1.1) the progenitor entered the virial radius of the host close to the upper bound of our preferred accretion time-window ($t_{\mathrm{acc}} \in 4-8$ lookback Gyr) with a very low angular momentum orbit, 
leading to immediate stripping (e.g., $t_\mathrm{strip} \gtrsim 4$ lookback Gyr) and shells that phase-mixed by $z=0\,$; 
or (1.2) the host is in a rich environment and undergoes many mergers over the next few Gyr (or a sufficiently disrupting major merger), speeding up the phase mixing of the shells created by this progenitor.

2) Some of the progenitors are harassed by other massive satellites of the central, due to the busy environment in the host halo.
Using movies, we identify this scenario taking place for some of the most massive centrals in our sample. 
We detect multiple ($\sim$3 -- 4) major mergers happening in quick succession. 
Typically, only one or two of these massive satellites manage to form shells, with the orbits of the other satellites arriving at similar times being strongly deviated. As a result, some of these progenitors are still orbiting the central host at $z=0$, while others have been stripped on less radial orbits and thus failed to form shells.

\subsection{Tracing the Evolution of a Shell Galaxy}
\label{subsec:mergerTree}
\begin{figure*}
\vspace{-0em}\includegraphics[width=.98\textwidth]{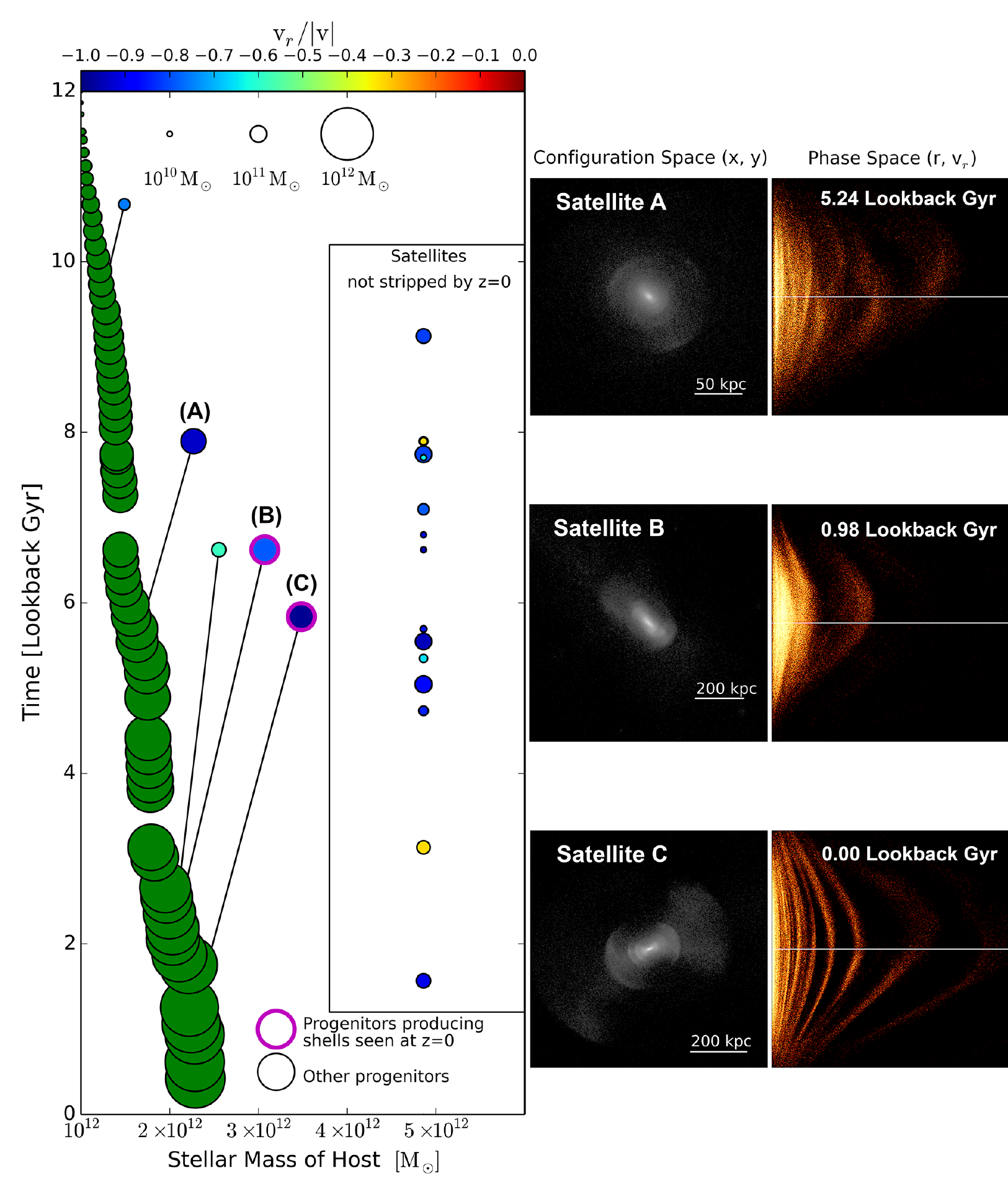}
\caption{ Time evolution of a shell galaxy in a cosmological setting. The left panel displays the merger tree of the final host (shown in green) at each time step, with satellites depicted on its right. The progenitors responsible for $z=0$ shells are marked by a magenta contour. All satellites of the host are represented in the figure based on their accretion time. Only five of the satellites have been stripped by $z=0$, with the black lines connecting them to their corresponding stripping time. All the other satellites in the middle column have not been stripped by $z=0$. The size of each circle indicates the total mass of the progenitor, while the color corresponds to the fraction of their velocity oriented along the radial direction, both measured at accretion time.
The right-most two columns show the entire halo in configuration and phase space at different times. The two shell-forming progenitors (satellites B and C) correspond to high-mass ratio radial mergers that took place in the relatively recent past (accretion times between $4-8$ Gyr ago). Moreover, we find that one of the high mass progenitors accreted earlier (satellite A) also forms shells, but those shells are short-lived and cannot be observed at $z=0$. }
\label{fig:mergerTree}
\end{figure*}

In order to get a better grasp of the time evolution of a shell galaxy, we turn now to studying one representative example of a galaxy with shells at redshift $z=0$. 
The merger tree of this galaxy is presented in the left panel of Figure \ref{fig:mergerTree}, with green circles marking the evolution of the host galaxy in time, and the size of the circle indicating the total mass of the galaxy. 
Throughout its history, the host accretes several smaller galaxies, marked with circles indicating their corresponding mass and their y-position marking the time of accretion of each satellite. 
Moreover, the color of each circle indicates how radial the orbit of the satellite was when entering the virial radius, as indicated by the colormap at the top of the panel. 
The darkest blue circles correspond to infall orbits with very low angular momentum (i.e., radial trajectories).
For those satellites that are stripped before $z=0$, 
a black line connects the accretion time to the green circle representing the host at the stripping time itself. 
Quite a few of the satellites do not get stripped by $z=0$ and they are included in the inset box in the middle of the figure. 

We mark the most massive three satellites accreted and stripped before $z=0$ with letters A through C, in order of their accretion times.
The shells we observe at $z=0$ were produced by satellites B and C, which are marked with magenta circles. These progenitors were accreted $\sim$6.6 and 5.8 Gyr ago and stripped $\sim$2.6 and 1.8 Gyr ago.
The corresponding merger stellar mass ratios of the two progenitors were $\mu_{\mathrm{stars,B}} = 0.60$ and $\mu_{\mathrm{stars,C}} = 0.47$. Their radial velocity ratios were $\left(\mathrm{v}_\mathrm{r}^{\mathrm{rel}}/|\mathrm{v}^{\mathrm{rel}} \right)_B = -0.79$ and \mbox{$\left(\mathrm{v}_\mathrm{r}^{\mathrm{rel}}/|\mathrm{v}^{\mathrm{rel}} \right)_C = -0.98$}. 
Thus, we see that both of these satellites correspond to roughly major mergers, with progenitors accreted on relatively radial orbits at an intermediate time in the past ($\sim$6 Gyr ago), in agreement with the results in Figure \ref{fig:threeDim}.

\begin{figure*}
\vspace{-0em}\includegraphics[width=1.02\textwidth]{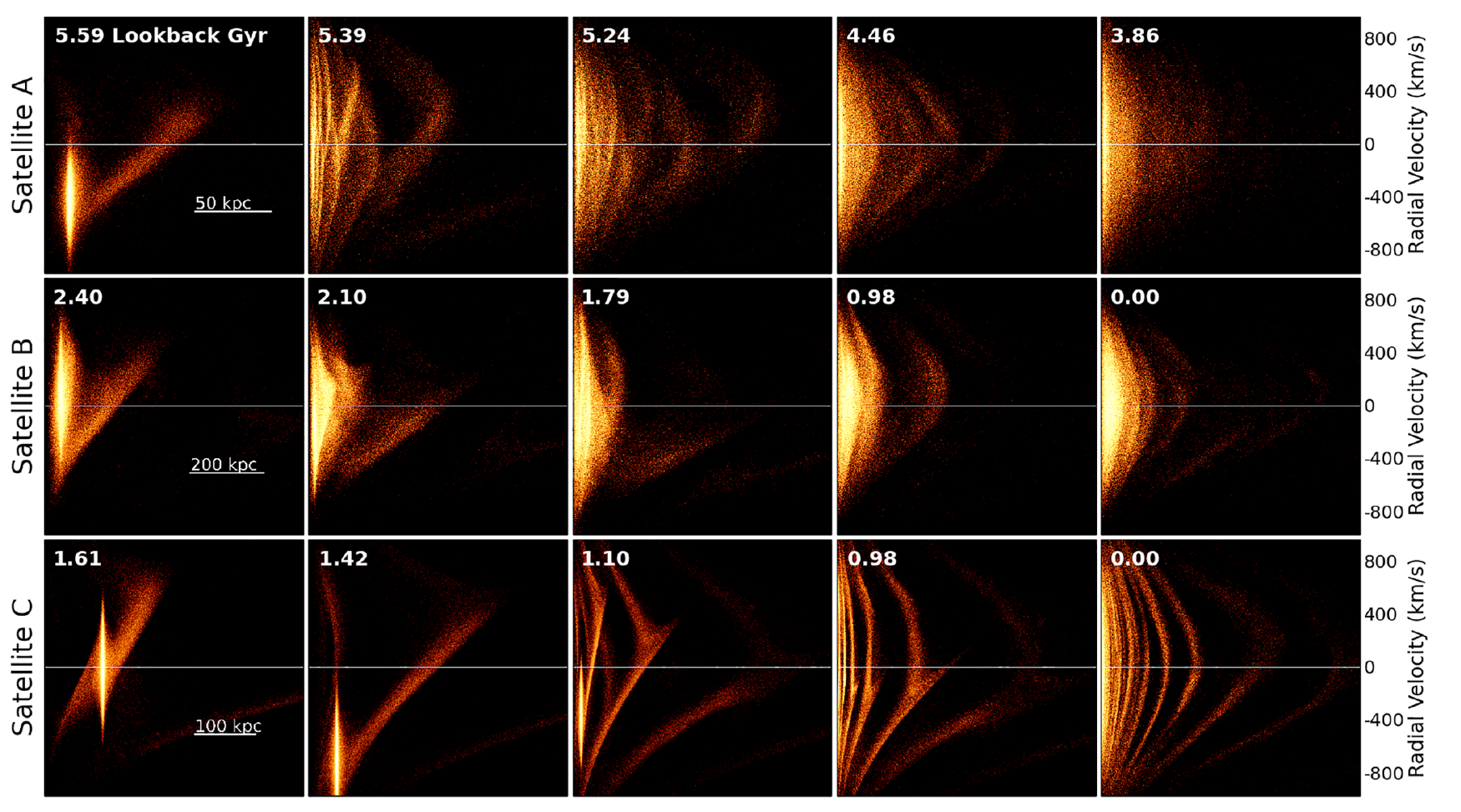}
\caption{Phase space ($v_r$ vs. $r$) evolution of the three satellites (A, B, C) from Figure~\ref{fig:mergerTree}. Each stamp includes at the top the time in lookback Gyr. Satellite A starts to be stripped almost 6 Gyr ago, it forms shells observable in the next $\sim$3 stamps, yet the shells phase mix rather quickly (within $\sim$1 Gyr). On the other hand, both satellites B and C form shells still visible at $z=0$ (see right-most stamps in rows 2 and 3). We note that the stars in satellite C were more dynamically cold (as marked by the smaller phase-space volume they occupy prior to being stripped). As a result, the shells formed by satellite C are less diffuse than those formed by satellite B.}
\label{fig:PhaseEvolution}
\end{figure*}

A closer investigation of this galaxy reveals that a third satellite (A) formed shells at an earlier time in the host's past. 
This satelllite was accreted $\sim$8 Gyr ago and stripped $\sim$6 Gyr ago.
Due to the smaller mass of the host when satellite A was stripped, this first set of shells forms closer to the core of the galaxy, 
leading to a shorter phase-mixing timescale. 
Figure~\ref{fig:PhaseEvolution} presents the phase-space evolution of the stars accreted from each of the three satellites. 
The shells formed by satellite A are relatively short-lived and they phase-mix in \mbox{$\sim$1 -- 1.5 Gyr} - before the next two shell-forming progenitors start to be stripped by the main host. On the other hand, the shells formed by satellite B are still visible at $z=0$, more than 2 Gyr after the satellite has been stripped. 
Comparing the phase-space evolution of satellites B and C, 
we find that satellite C appears colder than satellite B right before stripping (left column in Figure~\ref{fig:PhaseEvolution}). As a result, the shells formed by satellite C are less diffuse than those formed by satellite B.

\section{Varied Cosmological Scenarios for the Formation of Shell Galaxies}
\label{sec:studyCases}

Using a full cosmological simulation such as Illustris presents several advantages when studying the incidence and formation of stellar substructures such as shells. The previous sections relied on the good statistics offered by a cosmological volume in order to investigate the most common merger events producing shells and the incidence of these structures across different mass and redshift bins. Yet another advantage of our simulations compared to idealized mergers is that we naturally come across special shell-forming scenarios that we expect to also occur in the real Universe. This section is dedicated to investigating such special cases, some of which correspond to shell-forming satellites that lie outside the preferred region of parameter space in Figure~\ref{fig:threeDim}. In other cases, progenitors with high mass ratios arriving at intermediate times and on low angular momentum orbits can fail to produce shells due to special conditions such as those outlined below. The examples provided in this section help create a more complete picture of shell formation.

\subsection{Shell Galaxies with Multiple Shell-Forming Progenitors}
\label{subsec:multipleProg}
As already exemplified by Figures~\ref{fig:mergerTree} and \ref{fig:PhaseEvolution}, multiple satellites accreted on different orbits and at different times can contribute to separate stellar shell structures observed at $z=0$.
Out of the $39$ shell galaxies in our sample, 10 galaxies have two different progenitors contributing to redshift $z=0$ shells. All these ten host galaxies have $z=0$ masses $\mathrm{M}_{200\mathrm{crit}} \gtrsim 1.2 \times 10^{13}\, \mathrm{M}_\odot $, placing them in top $\sim$50\% of our sample of galaxies ordered by mass. Moreover, in about 5 of these galaxies we can detect faint traces of phase-mixed shells from another 1-2 progenitors. Similar to the case of satellite A in Figure~\ref{fig:mergerTree}, these progenitors were accreted at early times ($t_{\mathrm{acc}} \gtrsim 8$ Lookback Gyr), their stars were stripped and deposited rather close to the center of the host (relative to the extent of the stellar halo at $z=0$), and the shells phase-mixed within a few Gyrs. 

\begin{figure*}
\vspace{-0em}\includegraphics[width=\textwidth]{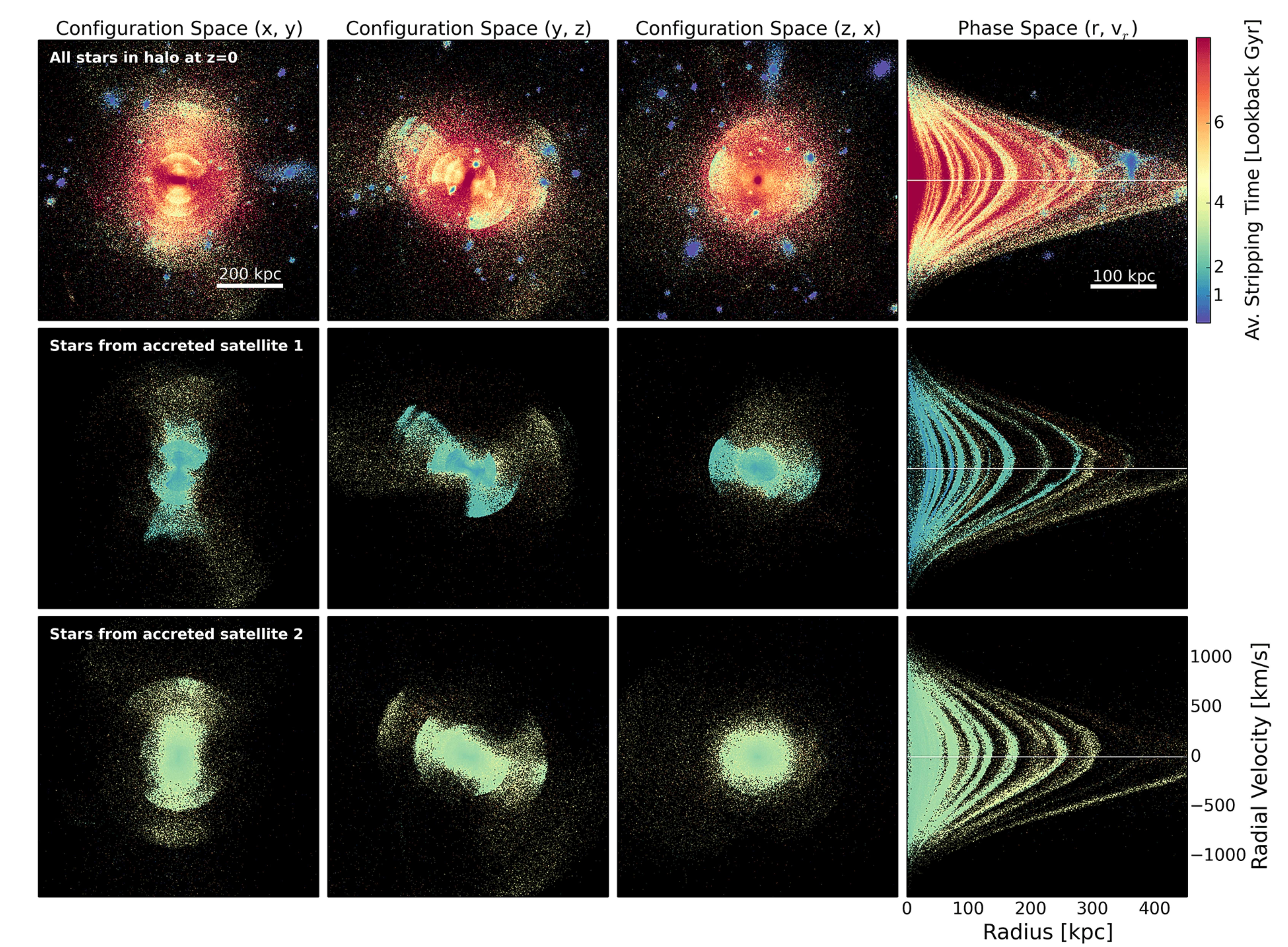}
\caption{Example of a redshift zero galaxy with $\mathrm{M}_{200\mathrm{crit}}(z=0) = 4.6 \times 10^{13}\, \mathrm{M}_\odot$ and $\mathrm{M}_{\mathrm{stars,tot}}(z=0) = 1.3 \times 10^{12}\, \mathrm{M}_\odot$ that had two different shell-forming progenitors responsible for the shells we see at $z=0$ (top row showing the entire halo). The first three columns correspond to configuration space in 3 different projections, while the last column shows the phase space ($\mathrm{v}_\mathrm{r}$ vs. $\mathrm{r}$) distribution of all stars. Each star in the image is colored based on its stripping time in units of lookback Gyr. We identify two different progenitors responsible for the shells at z=0, and the stars that belonged to each one of these satellites are shown in their current $z=0$ configuration in the second and third rows, respectively. The two shell-forming progenitors were accreted $\sim$6.5 Gyr ago and they were stripped $\sim$2 -- 3 Gyr ago. Their stellar mass ratios are $\mu_{\mathrm{stars}}(t_\mathrm{acc}) = 0.15$ and $0.24$, and they both arrived on very radial orbits with $v_r/|v|= -0.99$ and $-0.84$.}
\label{fig:twoProgenitors}
\end{figure*}

Depending on the exact stripping times and orbits of each shell-forming progenitor, the resulting shells from two independent satellites can appear more or less aligned or lie in a similar plane. One remarkable example of numerous shells formed by two separate progenitors is shown in Figure \ref{fig:twoProgenitors}. We count at least 12 shells at galactocentric distances $>$100 kpc, with several other shells visible at smaller radii. 
The host galaxy has a total mass of $\mathrm{M}_{200\mathrm{crit}}(z=0) = 4.6 \times 10^{13}\, \mathrm{M}_\odot$ and total stellar mass of $\mathrm{M}_{\mathrm{stars,tot}}(z=0) = 1.3 \times 10^{12}\, \mathrm{M}_\odot$. The two shell-forming progenitors have relatively small stellar mass ratios compared to our complete sample of shell-forming satellites ($\mu_{\mathrm{stars}}(t_\mathrm{acc}) = 0.15$ and $0.24$), but this is compensated by the very radial infall orbits ($v_r/|v|= -0.99$ and $-0.84$, respectively). These values place the two satellites in the lower left corner of the triangle in Figure~\ref{fig:threeDim}. 
The first shell-forming progenitor passes close to the center of the host 6 different times before getting completely stripped.
In these final phases, its orbit is disturbed by a massive satellite of the host, which causes the misaligned shells (most visible in the (y,z) projection).
The second satellite has a less eventful history, and it produces shells aligned with its infall trajectory, whilst the core of the satellite survives 3 different pericenter passages.

\begin{figure*}
\vspace{-0em}\includegraphics[width=\textwidth]{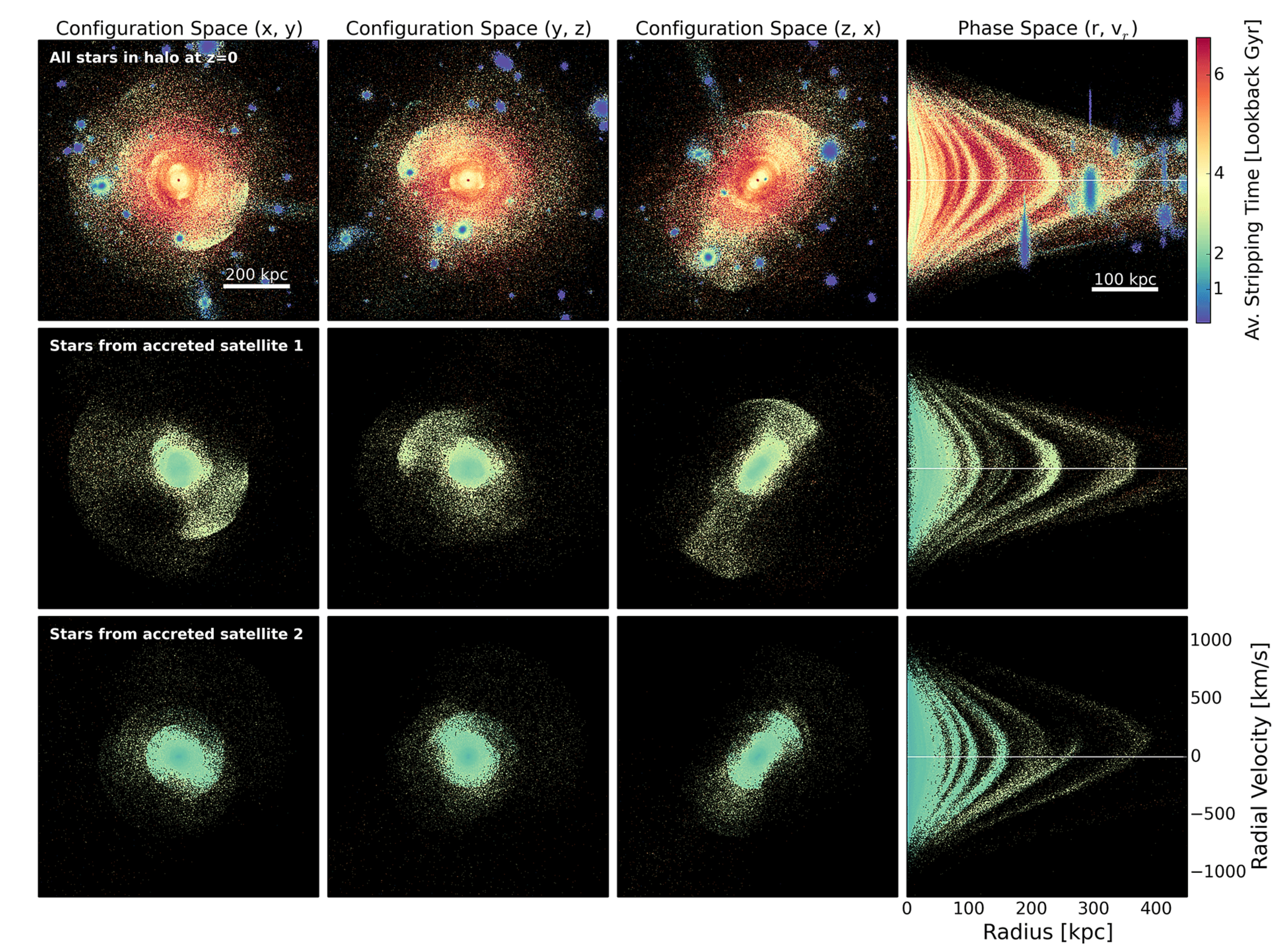}
\caption{Example of a redshift zero galaxy with $\mathrm{M}_{200\mathrm{crit}}(z=0) = 4.5 \times 10^{13}\, \mathrm{M}_\odot$ and $\mathrm{M}_{\mathrm{stars,tot}}(z=0) = 1.2 \times 10^{12} \, \mathrm{M}_\odot$ that had two different shell-forming progenitors responsible for the shells we see at $z=0$ (top row showing the entire halo). The first three columns correspond to configuration space in 3 different projections, while the last column shows the phase space ($\mathrm{v}_\mathrm{r}$ vs. $\mathrm{r}$) distribution of all stars. Each star in the image is colored based on its stripping time in units of lookback Gyr. We identify two different progenitors responsible for the shells at z=0, and the stars that belonged to each one of these satellites are shown in their current $z=0$ configuration in the second and third rows, respectively. The two shell-forming progenitors were accreted $\sim$7.3 and $5.9$ Gyr ago, respectively, and they were stripped $\sim$2.1/1.6 Gyr ago. Their stellar mass ratios correspond to roughly 1:2 mergers with $\mu_{\mathrm{stars}}(t_\mathrm{acc}) = 0.50$ and $0.51$, and they both arrived on very radial orbits with $v_r/|v|= -0.84$ and $-0.86$. }
\label{fig:twooProgenitors}
\end{figure*}

Another example of a shell galaxy with two separate shell-forming progenitors is presented in Figure~\ref{fig:twooProgenitors}. The top row shows all the stars in the halo with $\mathrm{M}_{200\mathrm{crit}}(z=0) = 4.5 \times 10^{13}\, \mathrm{M}_\odot$ and $\mathrm{M}_{\mathrm{stars,tot}}(z=0) = 1.2 \times 10^{12} \, \mathrm{M}_\odot$.
The first satellite was accreted/stripped 7.3/2.1 Gyr ago, while the second one was accreted 5.9 Gyr ago and stripped 1.6 Gyr ago. 
In this case, both mergers producing shells correspond to major merger events with $\mu_{\mathrm{stars}}(t_\mathrm{acc}) = 0.50$ and $0.51$, as well as very radial orbits ($v_r/|v|= -0.84$ and $-0.86$, respectively). Both satellites produce shells visible in all three projections, and the ensemble of shells formed by the two progenitors accounts for all the shells observed at $z=0$ (compare the three phase-space images in the last column of Figure~\ref{fig:twooProgenitors}).

\subsection{Satellite of a Satellite Forming Shells}
\label{subsec:satofsat}

\begin{figure*}
\vspace{-0em}\includegraphics[width=\textwidth]{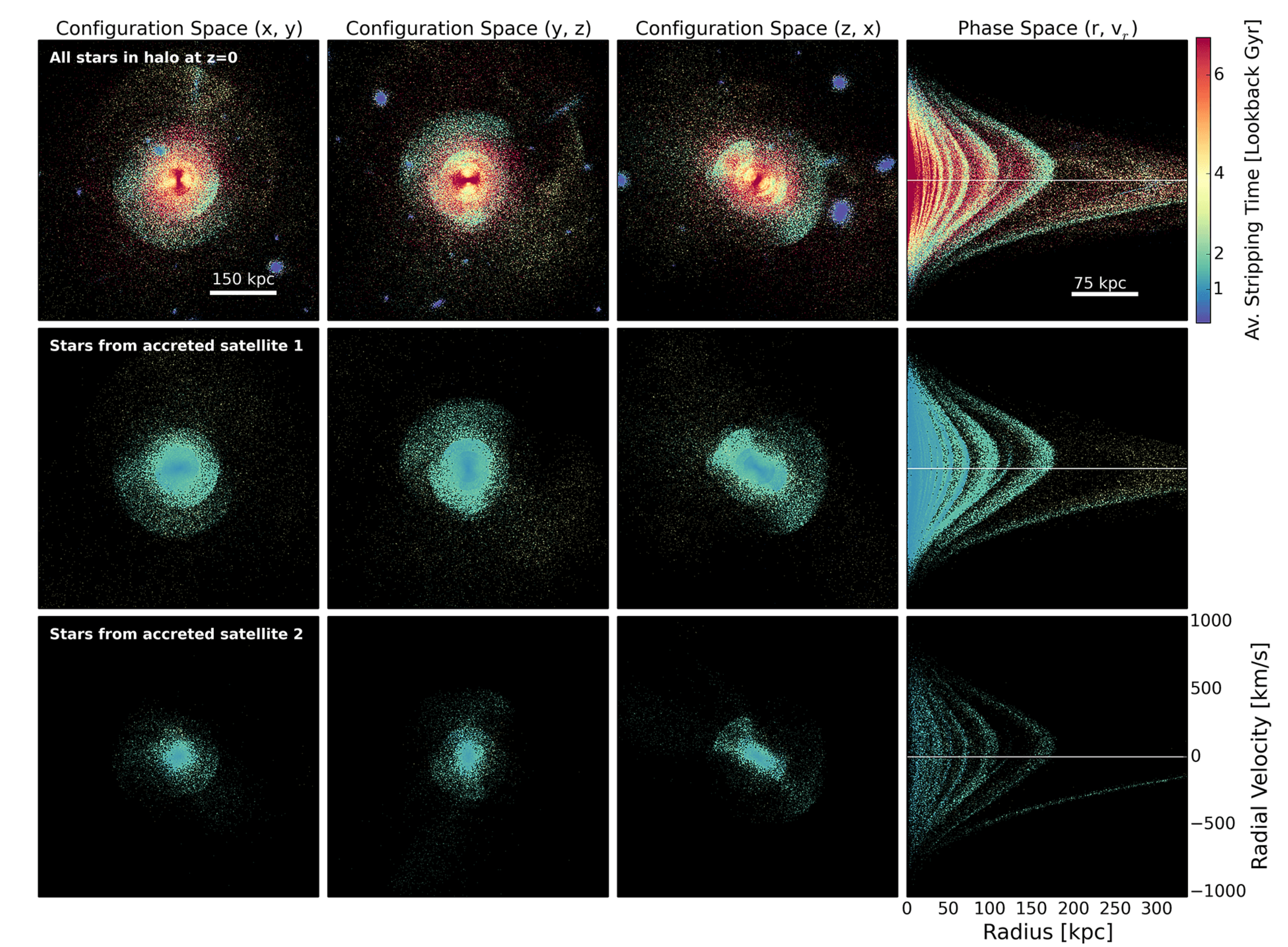}
\caption{Example of a redshift zero galaxy (top row) with shells formed by a massive progenitor (middle row) and its smaller companion (bottom row). The host galaxy has $\mathrm{M}_{200\mathrm{crit}}(z=0) = 1.9 \times 10^{13}\, \mathrm{M}_\odot$ and $\mathrm{M}_{\mathrm{stars,tot}}(z=0) = 5.2 \times 10^{11} \,\mathrm{M}_\odot$, and the top row shows all the stars in the halo at $z=0$, colored based on their individual stripping times. Stars stripped a very long time ago (as well as in-situ stars that formed early) are shown in red, stars from more recently stripped satellites are colored in yellow/blue, and the satellites in the top row that have yet to be stripped by $z=0$ are shown in purple. The first three columns correspond to configuration space in 3 different projections, while the last column shows the phase space ($\mathrm{v}_\mathrm{r}$ vs. $\mathrm{r}$) distribution of all stars. Redshift $z=0$ shells are composed of stars stripped from a massive progenitor (middle row) and its smaller companion or satellite-of-a-satellite (bottom row). The two satellites have been accreted at the same time ($t_{\mathrm{acc}} = 4.3$ lookback Gyr), and their stars have been stripped about $1.1-1.6$ Gyr ago. The stellar mass ratios of the two satellites are \mbox{$\mu_{\mathrm{stars}}(t_\mathrm{acc}) = 0.75$} and $0.066$, respectively; and the radial velocity ratio of the more massive satellite (which dictates the infall orbit of both galaxies) is $v_r/|v| = -0.81$.}
\label{fig:satelliteofSatellite}
\end{figure*}

Unlike the examples shown previously, where the two shell-forming progenitors were independent galaxies following their separate orbits, we also find cases in which the shell-forming progenitors are physically associated. 
One such interesting example corresponds to a satellite falling in the potential well of the main host together with its smaller companion (see Figure~\ref{fig:satelliteofSatellite}). While the smaller companion (or satellite-of-a-satellite) has a much lower stellar mass ratio \mbox{$\mu_{\mathrm{stars}}(t_\mathrm{acc}) = \mathrm{M}_{\mathrm{satellite-of-satellite}}/\mathrm{M}_{\mathrm{host}} = 0.066$} than the satellite to which it is gravitationally bound \mbox{($\mu_{\mathrm{stars}}(t_\mathrm{acc}) = \mathrm{M}_{\mathrm{satellite}}/\mathrm{M}_{\mathrm{host}} = 0.75$)}, it still succeeds to produce $z=0$ shells by tagging along the orbit of the bigger progenitor. The stars that used to belong to the satellite-of-a-satellite are shown in the bottom row of Figure~\ref{fig:satelliteofSatellite}, while the middle row shows stars belonging to the main shell-forming progenitor. By comparing the phase-space panel for both satellites, we find that both progenitors trace the same shells, with the caveat that the smaller companion provides significantly ($\sim$11$\times$) fewer stars. By tracking the trajectory of the two satellites, as well as comparing their corresponding $z=0$ phase-space stamps, we establish that the smaller companion follows the same orbit as the bigger satellite, and its dynamical properties throughout its final infall stages are entirely dictated by the gravitational interaction between the main host and the more massive satellite. We thus only include the more massive satellite and exclude the case of the satellite-of-a-satellite from our results in Figure~\ref{fig:threeDim}. This represents an atypical case in which a merger event with a relatively small stellar mass ratio can produce shells in $z=0$ massive ($\mathrm{M}_{200\mathrm{crit}} > 6\times10^{12}\, \mathrm{M}_\odot$) galaxies. The ensemble of the two satellites in Figure~\ref{fig:satelliteofSatellite} (massive satellite + smaller companion) essentially act together like a single satellite of combined stellar mass ratio $\mu_{\mathrm{stars}}(t_\mathrm{acc}) = (\mathrm{M}_{\mathrm{satellite}}+ \mathrm{M}_{\mathrm{satellite-of-satellite}}) /\mathrm{M}_{\mathrm{host}} = 0.82$, entering the virial radius of the main host about 4.3 Gyr ago, with a radial velocity ratio $v_r/|v|(t_{\mathrm{acc}}) = -0.81$. Comparing these values to the expected values for a shell-forming progenitor in Figure~\ref{fig:threeDim}, this combined system would lie in the upper-central region of the preferred red triangle in the ($v_r/|v|$, $\mu_{\mathrm{stars}}$) space. 

We note that the satellite-of-a-satellite in the bottom panel of Figure~\ref{fig:satelliteofSatellite} is the smallest shell-forming progenitor ($\mathrm{M}_{\mathrm{stars,tot}}(t_{\mathrm{acc}}) = 1.7 \times 10^{10} \,\mathrm{M}_\odot$, $\mathrm{M}_{\mathrm{tot}}(t_{\mathrm{acc}}) = 5.2 \times 10^{10} \,\mathrm{M}_\odot$) in our sample, with the next smallest shell-forming progenitor having a total mass more than 4 times higher than this satellite-of-a-satellite. Nonetheless, even for this lowest mass case, we can still clearly distinguish shell-like structures. We discuss the resolution limits of our study in Appendix~\ref{sec:resolution}.

\section{Discussion}
\label{sec:discussion}

\subsection{Comparing the Fraction of Shell Galaxies with Observations}
\label{subsec:discussFraction}

As mentioned in the Introduction, a consensus has yet to be reached on the observed fraction of shell galaxies, with
different groups reporting measurements ranging from $22\%$ \citep{Taletal2009} to $3.5\%$ \citep{Krajnovicetal2011}. 
The disparity may be caused, in part, by variations in the mass and color of the galaxies in the sample:
shells are more common in red galaxies ($14\%$ exhibiting shells) than in blue galaxies ($6\%$) \citep{Atkinsonetal2013}.
The incidence of shells is also affected by the environment: stellar shells are found more frequently around isolated galaxies, while the fraction of shells drops considerably in clusters or rich groups: $48\%$ of isolated galaxies have \mbox{shells vs.} $3.6\%$ in rich environments \citep[e.g.,][]{Malin&Carter1983, Reduzzietal1996, Colbertetal2001}.
Lastly, the measured fraction of shells may also be affected by the surface brightness limits of each survey, 
as well as by the techniques used to identify and enhance the shells. 
Fortunately, \mbox{these uncertainties could} \mbox{be soon} \mbox{resolved thanks} to \mbox{the significant} recent effort in \mbox{low surface brightness} observations \citep[e.g.,][]{Janowieckietal2010, Miskolczietal2011, Atkinsonetal2013, MartinezDelgadoetal2010, Abraham&vanDokkum2014, Ducetal2014, Trujillo&Fliri2016}.

The incidence of shells in our sample of massive early-type galaxies is in overall agreement with current observational limits. 
Nonetheless, the fraction of shell galaxies we present in Figure \ref{fig:massDistofShells} is somewhat higher than the fractions observed in two recent surveys \citep{Krajnovicetal2011, Duc2016}, 
which investigate similarly sized samples ($\sim$200 early-type galaxies).
This effect could be due to 3 main factors related to the way we identify the shell galaxies: 1) we do not apply a surface brightness cut, but instead, we place ourselves in the best position to detect all shell galaxies in our sample of simulated galaxies, 
in order to have a complete sample of shells at $z=0$ in Illustris and to get a statistically accurate understanding of the dominant formation scenarios (see Section~\ref{sec:progenitors} on the progenitors of shell galaxies);
2) we show that for massive ellipticals, the fraction of galaxies with shells decreases mildly with increasing redshift (Figure \ref{fig:massDistofShells}),
which might cause surveys that include higher-redshift galaxies to report slightly lower fractions of shell galaxies;
3) we can more easily identify shells by studying multiple projections in our simulation, while observations are limited to studying a single projection normal to the line of sight.
Our estimates indicate that restricting to a single projection causes a reduction in the measured fraction of shells of $f_{\mathrm{obs}} \sim 0.8 f_{\mathrm{sim}}$ (see Appendix~\ref{sec:projectionEffects} for more details). For our full sample of $z=0$ galaxies with $\mathrm{M}_{\mathrm{200crit}} > 6 \times 10^{12} \mathrm{M}_\odot$, the measured fraction of shells would drop to $14\% \pm 3\%$.

\subsection{Mass Distribution of Shell Galaxies}
\label{subsec:discussMass}

In \S\ref{subsec:fraction} and \S\ref{subsec:redshift}, we find that the fraction of shell galaxies increases when we limit our sample to the most massive galaxies in Illustris, 
and that it decreases with increasing redshift. 
Moreover, in Section~\ref{sec:progenitors}, we show that shells visible at $z=0$ form through mergers with relatively massive satellites (stellar mass ratios between $\sim$1/10 and $1$), accreted approximately between 4 and 8 Gyr ago. 
Mergers with relatively high total mass ratios occur more frequently for high descendant masses and they are more common at higher redshifts \citep[e.g.,][]{Rodriguez-Gomezetal2015}. 
Therefore, in agreement with Figure \ref{fig:massDistofShells}, we expect the highest mass galaxies at $z=0$ to have had slightly more major mergers in their recent past, making them more likely to have shells visible today.
We note here that the number of shell galaxies in the most massive bins in Figure \ref{fig:massDistofShells} represents in fact a lower limit. These galaxies lie at the centers of rich galaxy groups,
which makes it more difficult to detect shells 
due to the presence of numerous satellites and a multitude of other low surface brightness features. 
This could provide a hint as to why we see the supremum distance between the CDFs in the insets in Figure \ref{fig:massDistofShells} at intermediate masses $\mathrm{M}_{\mathrm{200vir}} \in 2-6 \times 10^{13} \,M_{\odot}$ in our sample. 
We cannot, however, exclude that this indeed a physical effect, and that the very highest mass galaxies at the centers of clusters or rich groups experience competing effects that drive the fraction of shells down. 
According to previous observational surveys, the incidence of shells is higher for galaxies in the field vs. for galaxies in rich groups and clusters \citep{Malin&Carter1983, Colbertetal2001, Reduzzietal1996}.
The lower fraction of shells in rich environments might be related to the intense recent merging activity, possibly implying faster phase mixing times and shorter lifetimes for any shell structure. 

\subsection{Redshift Evolution of the Fraction of Shells}
\label{subsec:discussRedshift}

As mentioned above, major mergers are more common for more massive descendants, and more frequent at higher redshifts \citep{Rodriguez-Gomezetal2015}. 
As a result, we would expect stellar shells to form more often in the past. 
However, Figure~\ref{fig:redshiftEvolution} shows that the fraction of massive ellipticals with shells is mildly decreasing with redshift.
It should be remembered that Figure~\ref{fig:redshiftEvolution} does not quantify the fraction of progenitors that are successful at forming shells.
Instead, the observed trend could result from the shorter lifetimes of high-$z$ shells. 
Higher redshift implies smaller hosts and shorter dynamical timescales, which in turn lead to the faster phase mixing of shells. 
Moreover, shell material is deposited closer to the center of the host, making it even harder to detect them. 
While shells visible at $z=0$ seem to have survived on average $\sim$2 -- 4 Gyr (see stripping time panel in Figure \ref{fig:progenitors} and discussion in \S \ref{sec:progenitors}), in the example we later discuss in Figure \ref{fig:mergerTree}, we find shells at $z \sim 0.5$ that phase mix within only $\sim$1 Gyr. These shorter phase mixing times can provide a possible explanation for the decrease in the fraction of massive galaxies with shells at higher redshifts (Figure \ref{fig:redshiftEvolution}).

\subsection{Progenitors of Shells}
\label{subsec:dicussProgenitors}

Prior to this study, the most commonly invoked model for shell formation involved radial minor mergers. \cite{Quinn1984} was the first to suggest that shells could be remnants of a small galaxy accreted by a much more massive primary host. Many of the simulations and semi-analytical models that followed have focused on idealized minor mergers, with satellites initialized on purely or nearly radial infall trajectories \citep[e.g.,][]{Dupraz&Combes1986,  Hernquist&Quinn1987b, Hernquist&Quinn1988, Hernquist&Quinn1989,  Seguin&Dupraz1996, Weil&Hernquist1993, Sanderson&Bertschinger2010, Bartoskovaetal2011, Ebrova2012, Ebrova2013}. Moreover, in most cases, these set-ups correspond to direct hits where the satellite is initialized with $\sim$zero impact parameter.
Major mergers (mass ratios $\mu \gtrsim$ 1:10) have also been explored, although they have mainly been considered as special cases.  \cite{Hernquist&Spergel1992} simulated the formation of shells in a prograde collision of two disk galaxies of equal mass \citep[see also][]{GonzalezGarcia&Balcells2005}. \cite{GonzalezGarcia&vanAlbada2005a, GonzalezGarcia&vanAlbada2005b} present simulations of mergers between two elliptical galaxies (with and without a dark halo, $\mu$ of 1:1, 1:2, 1:4) and find shells formed through phase wrapping of stars from the accreted satellite, just like in previous simulations of minor mergers. The shells they find are sharper when a dark halo is included. 
More recently, \cite{Cooperetal2010} reported shells in more than half of the 6 DM halos with $\sim$Milky Way masses simulated as part of the Aquarius project \citep{Springeletal2008}. 
Among these, three have relatively high numbers of significant progenitors (i.e., the host accreted many satellites of similar masses), while in a fourth case, the progenitor responsible for the shells contributes $95\%$ of the ex-situ material, corresponding to a mass greater than the Small Magellanic Cloud and a merger total mass ratio of 1:3 \citep{Cooperetal2011}.

Several observations of shell galaxies have indicated a possible major merger origin of the shells \citep{Schiminovichetal1995, Balcellsetal2001, Goudfrooijetal2001, Serraetal2006}. \cite{Goudfrooijetal2001} suggest that Fornax A (NGC 1316) hosted a major merger 3 Gyr ago based on the age of a bright subpopulation of globular clusters, as well as the presence of molecular gas at \mbox{$\sim$30 -- 50} arcsec from the galaxy center \citep{Sage&Galletta1993}.
It has also been suggested that metallicity gradients of the shell galaxies could help distinguish between major/minor mergers in the history of the host. 
As shown by \cite{Cooketal2016}, Illustris galaxies with higher ex-situ fractions have flatter metallicities profiles. \cite{Popetal2017} identified several shell galaxies in Illustris that underwent major mergers and exhibit outer shells more metal-rich than the surrounding stellar material in the host halo. This indicates that metallicity (or color) measurements can be used to probe the mass ratios of the mergers producing shells.
Recently, \cite{Carlstenetal2016} analyzed long slit spectra of 9 shell galaxies and found relatively shallow metallicity gradients, possibly indicating that these galaxies have undergone major mergers \citep[e.g.,][]{Kobayashi2004}. 

The results presented in \S\ref{subsec:recipe} and Figure~\ref{fig:threeDim} indicate that the order-zero recipe for shell formation requires satellites to have relatively radial infall orbits, high stellar mass ratios (roughly major mergers), and intermediate accretion times ($\sim$4 -- 8 Gyr ago). 
We use logistic regression (see Appendix~\ref{sec:logisticRegression}) to identify the region in the parameter space ($v_r, \mu_{\mathrm{stars}}$) where shell-forming progenitors are most likely to be found. Our results indicate that satellites involved in major mergers can have a much wider spread in the orientation of their trajectories at accretion time. On the other hand, less massive satellites need to have almost purely radial orbits when entering the virial radius of the host in order to produce shells.
We do not exclude that satellites with smaller stellar mass ratios ($\mu_{\mathrm{stars}} <$1:10) could form shells, but we find that this requires their infall trajectories to be very fine-tuned -- i.e., almost perfectly radial -- which makes these events less frequent. 

In an early analytic study of the effects of dynamical friction on the radial distribution of shells around ellipticals, \cite{Dupraz&Combes1987} found a somewhat similar trend: the more massive the satellite, the larger the range of impact parameters allowed for the formation of shells. To reach this conclusion, \cite{Dupraz&Combes1987} required the cores of the accreted satellites to survive several pericenter approaches, in order to explain observations of inner shells \citep[e.g.,][]{Wilkinsonetal1987}. It is expected that new generations of shells are produced as more and more stars are stripped from the companion at each successive passage near the central galaxy \citep{Seguin&Dupraz1996, Bartoskovaetal2011, Cooperetal2011, Ebrova2013}. This is in agreement with shell galaxies in Illustris, with studies of the satellites' trajectories indicating that the disruption of the companion happens over several pericentric approaches (see Figure~\ref{fig:projectionsColormaps}).
We have also checked that, on average, the first pericenter approach for the systems in our sample occurs at tens of kpc from the potential center of the host galaxy, namely in regions were the simulated host potentials are resolved with $\sim 10^5-10^6$ resolution elements, and hence 
the efficiency of tidal stripping is numerically properly captured.
More recently, \cite{BoylanKolchinetal2013} have studied the timescales and properties of mergers between $\Lambda$CDM halos, and \cite{Amorisco2017} has explored the role of dynamical friction in shaping how accreted satellites deposit their stars onto the host galaxy. The latter work shows that the efficiency of dynamical friction increases with the merger mass ratio, and that, crucially to the present work, the orbits of satellites with high $\mu$ (close to major mergers) are quickly radialized. As a consequence, in our simulation we find that a larger fraction of massive satellites have orbits that are radial enough to produce several generations of shells in the final stages of the merger, despite higher values of the orbital angular momentum at accretion. In addition to this driving effect, cosmological N-body simulations also indicate that more massive satellites have a mild bias for more eccentric infall orbits \citep{Tormen1997, Benson2005, Wangetal2005, Khochfar&Burkert2006, Jiangetal2015}.

In the current cosmological study, we consider massive galaxies that are predominantly found at the centers of groups and clusters. We argue that dynamical friction is an important ingredient in radializing the orbits of satellites involved in $\sim$major mergers, and thus it allows massive satellites to probe a wider range of initial impact parameters. Massive hosts will also suffer more major mergers in the last few Gyrs, making them more likely to host shells.
Moreover, interactions between very small mass ratio satellites with other incoming progenitors, as well as the constantly evolving host halo cause some of the small mass progenitors to be destroyed before they reach the center of the host. Cosmological simulations have the advantage of capturing the combined effects of dynamical friction, multiple mergers happening in quick succession, and the evolution of the host galaxy's halo. In this context, we find that close-to-major mergers are most likely to produce $z=0$ shell structures in massive hosts.

\section{Summary and Conclusions}
\label{sec:conclusions}

This paper provides the first study of the incidence and formation processes of shell galaxies in a cosmological setting, based on a large, statistically relevant sample of 220 massive ($\mathrm{M}_{200{\mathrm{crit}}} > 6 \times 10^{12}\, \mathrm{M}_\odot$) galaxies from the gravity+hydrodynamics Illustris simulation. We identify galaxies with shells in our sample using a two-step approach:

$\bullet$ \textit{Step 1:} We visually identify galaxies with shells using stellar surface density maps.

$\bullet$ \textit{Step 2:} We trace the history of all stars inside the $z=0$ halos. 

We use this approach to identify the individual stars and the progenitor galaxies responsible for forming shells, and we confirm the ex-situ origin of stellar shells in our simulation.
 
The shell galaxies we find in Illustris have many
varied morphologies that closely resemble observed images of shells. The simulation was not fine-tuned in any way to produce these tidal features; they are the natural result of galaxies' dynamics and assembly histories.
Based on our two identification steps, we find that 39 of the 220 galaxies in our sample exhibit shells at redshift $z=0$. This corresponds to a fraction of $18\% \pm 3\%$ shell galaxies. We expect observed fractions to be slightly lower than this due to 1) projection effects, 2) surface brightness limits, and 3) inclusion of higher redshift galaxies.
We find that the fraction of shells increases for higher mass cuts. Moreover,
for massive elliptical galaxies, the fraction of observable shell galaxies is lower at higher redshifts, possibly due to the shells' shorter lifetimes/faster phase-mixing timescales.
We also find that the mass distribution of galaxies with shells is relatively flat, with higher mass galaxies in the field having a slightly higher likelihood to form shells visible at $z=0$. 

A direct advantage of using a cosmological simulation such as Illustris is that we can study the type of mergers producing shells on a relatively large sample of galaxies, providing us with good statistics of varied assembly histories.
We find an order-zero recipe for satellites forming shells, based on the mass ratio, infall orbit and accretion time of the progenitor galaxies.
Our results indicate that  massive shell galaxies observed at $z=0$ form preferentially through relatively major merger ($\gtrsim$1:10 in stellar mass ratio). Shell-forming progenitors are accreted on low angular momentum orbits (i.e., radial infall trajectories), in a preferred time-window between $\sim$4 and 8 Gyr ago. Moreover, we find that satellites responsible for $z=0$ shell structures were stripped on average $\sim$2 Gyr ago, which suggests that shell phase-mixing times for $z=0$ halos should be comparable ($\sim$2 -- 3 Gyr).

We concentrate on a number of special cases of shell-forming progenitors that depart from the order-zero recipe of high mass mergers on radial orbits at intermediate accretion times, and that are a consequence of the additional complexity introduced by the cosmological setting. 
In our sample, 10 out of 39 shell galaxies have multiple progenitors that contribute to the $z=0$ shells, 
as well as
several additional satellites that formed shells in the past and phase-mix before $z=0$. An interesting and less common shell-forming scenario corresponds to progenitors that are being accompanied by their own smaller satellites (see Figure~\ref{fig:satelliteofSatellite}). Despite the low mass ratio of the satellite of a satellite, the smaller companion succeeds at forming $z=0$ shells
by tagging along the orbit of its parent galaxy. We also discuss how the busy environments of high-mass hosts can lead to massive satellites on radial orbits failing to form shells. 
Often times, the orbits of these progenitors can be significantly impacted by the presence of other satellites inside the halo of the central host.
As a result, some of the direct progenitors of massive galaxies that were initially accreted on low angular momentum orbits can ultimately fail to produce shells observable at $z=0$.

Overall, the results obtained in this paper indicate that major mergers account for a significant fraction of the shells observed in massive elliptical galaxies. We find that the key to forming shells lies in a very radial encounter between the satellite and the host galaxy in the final stages of the merger. Due to dynamical friction, more massive satellites are allowed to probe a wider range of impact parameters at accretion time, while small companions need almost purely radial infall trajectories in order to produce shells. 

The current study is devoted to a particular class of tidal features: stellar shells.
However, the methods developed in this paper can be applied to studying the incidence and formation mechanisms of a much wider range of low surface brightness features, encompassing the continuous range of morphologies connecting shells to umbrellas,
streams, and plumes. 
Similar studies will be of paramount importance in interpreting the large datasets that wide-area, low surface brightness surveys will soon provide.

\section*{Acknowledgments}
The authors thank the referee for helpful suggestions.
ARP would like to thank Vicente Rodriguez-Gomez for support on using merger trees in Illustris, 
Laura Sales for conversations on satellites' morphologies and rotational support,
Kathryn Johnston for stimulating discussions on the infalling orbits of the satellites,
Jieun Choi for useful comments on the draft,
Josh Speagle for discussions on logistic regression,
and Fernando Becerra for helpful comments on data visualization.

\appendix

\section{Projection Effects}
\label{sec:projectionEffects}

\begin{figure*}
\vspace{-0em}\includegraphics[width=.9\textwidth]{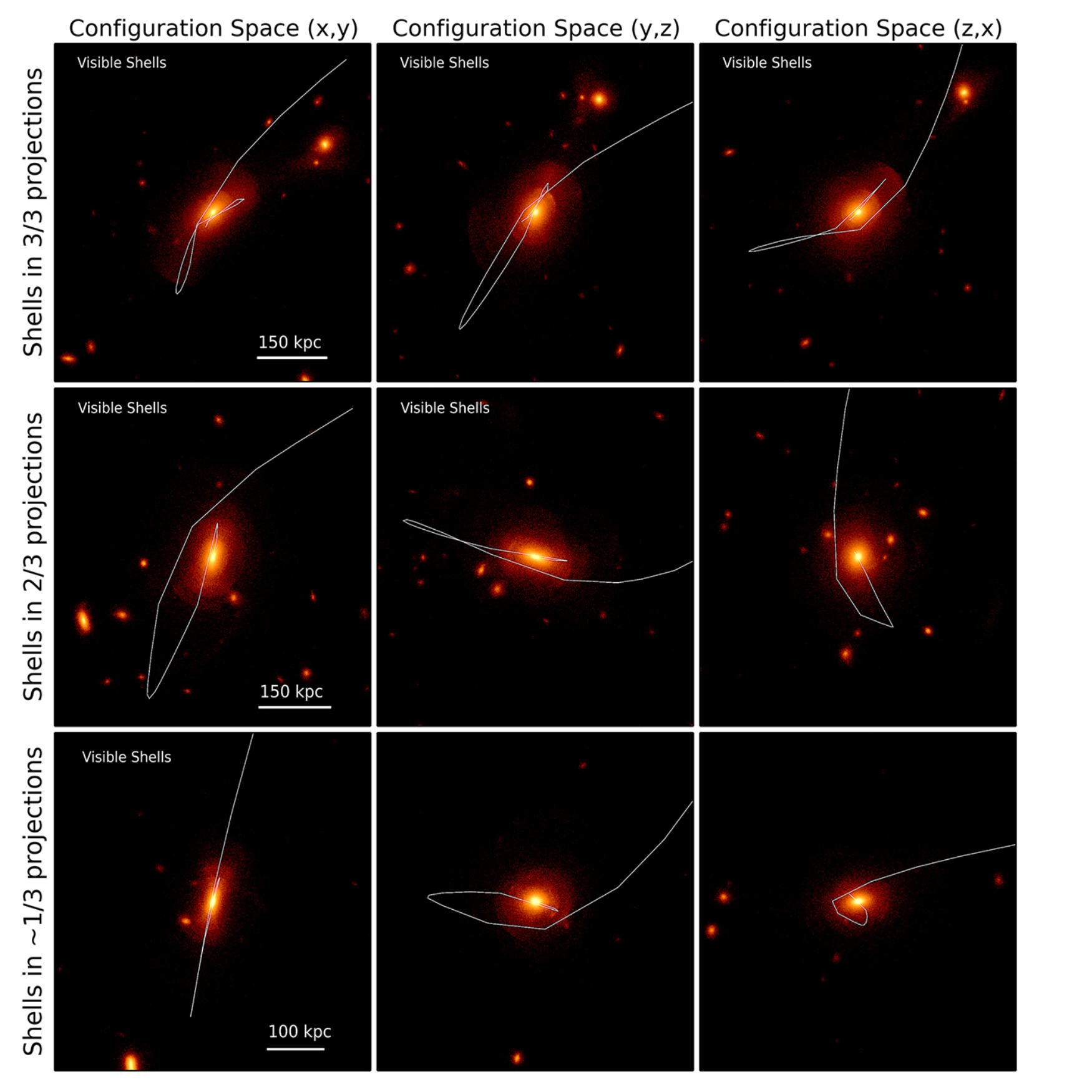}
\caption{
Examples of three different $z=0$ galaxies with shells visible in all three projections (top row), only 2/3 projections (second row), only 1/3 projections (bottom row). The infall trajectories of the shell-forming progenitors for each of the three galaxies are marked by white lines.}
\label{fig:projectionsColormaps}
\end{figure*}

\begin{figure*}
\vspace{-0em}\includegraphics[width=0.9\textwidth]{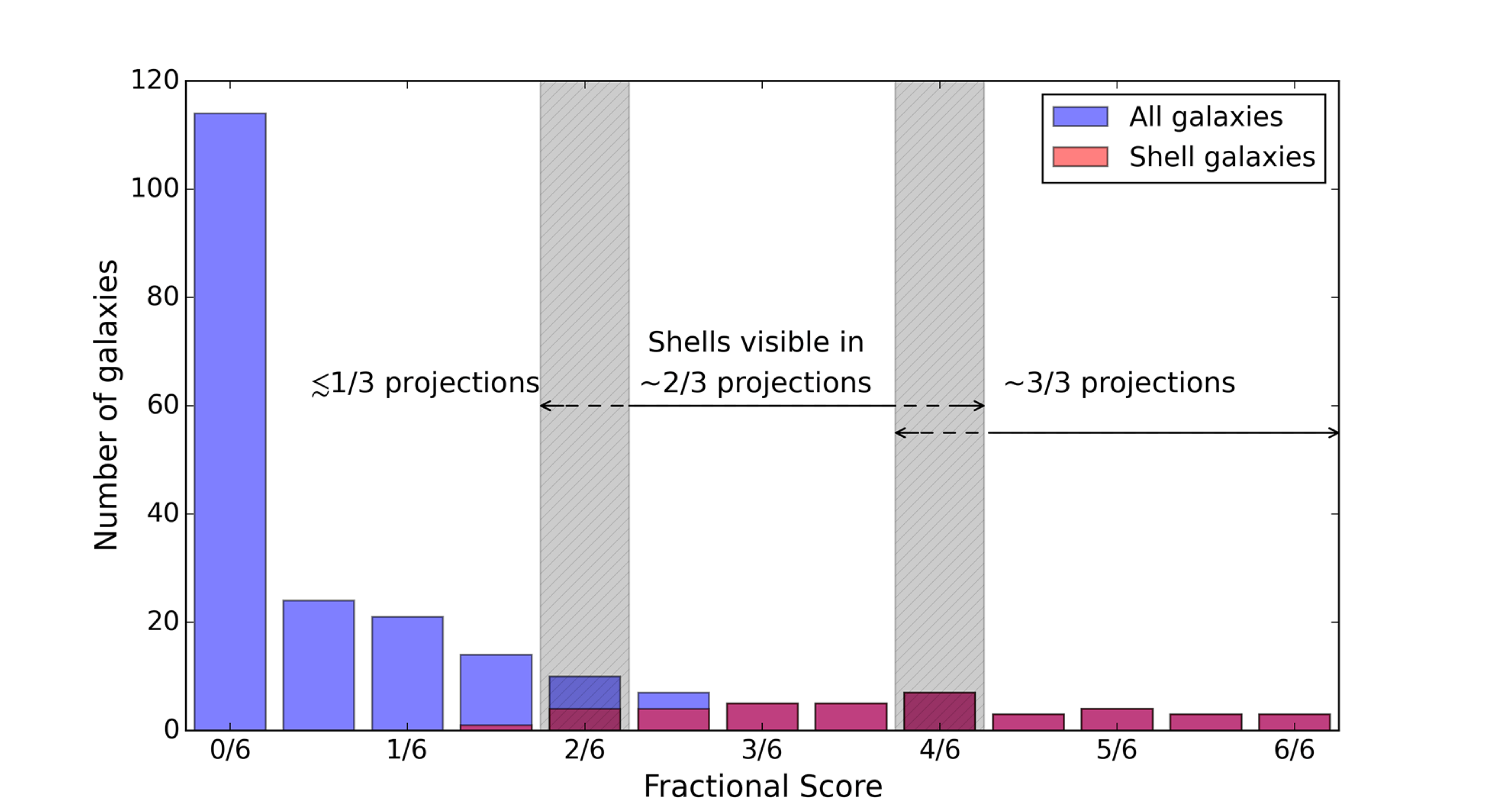}
\caption{Averaged score (between the three judges) for all the galaxies in our sample (shown in blue). Galaxies that have shells confirmed through both identification steps discussed in  \S\ref{subsec:galaxySample} and \ref{subsec:stellarHistoryCatalogs} are overplotted in red. We give a score of 2 (for 2, 3, or more well-defined shells), 1 (for 1-2 shell-like structures), or 0 (no shell detection) to each galaxy, as shown for example in Figure \ref{fig:scoring}. Since we score each different projection separately, each galaxy can have an averaged score between the judges ranging between 0 and 6. All galaxies with scores above $3/6$ are confirmed to have shells using stellar history catalogs, and the distribution of scores is relatively flat for shell galaxies.}
\label{fig:scoreDist}
\end{figure*}

As discussed in Section~\ref{subsec:fraction}, we expect our estimates for the fraction of shell galaxies in Figure~\ref{fig:massDistofShells} to be slightly higher than the observed fractions since we do not apply a surface brightness limit to the images we use to identify shells, we only include $z=0$ galaxies (while according to Figure~\ref{fig:redshiftEvolution}, at higher $z$, massive galaxies are less likely to exhibit shells), and lastly, we take advantage of all three projections. 
Some of the shells are only visible in certain projections depending on the infall trajectory of the shell-forming progenitor, and thus, observations limited only to one projection normal to the line of sight will underestimate the incidence of shell galaxies.

Figure \ref{fig:projectionsColormaps} shows three galaxies in our sample for which the shells are visible in $\sim$1, 2, or 3 projections, respectively. The infall trajectories of their corresponding shell-forming progenitors are overplotted with white lines.

In order to give a quantitative estimate for the extent to which observing shells in a single projection affects the observed fractions, we look at our distribution of shell scores shown in Figure \ref{fig:scoreDist}. Since we give scores of 2 (for 2, 3, or more well-defined shells), 1 (for 1-2 shell-like structures), or 0 (no shell detection), we expect shells visible in only one projection to have the smallest fractional score, followed by shells visible in 2, and finally, shells visible in 3 projections should have the highest scores.
To estimate this effect more precisely, we sum up the scores given to each galaxy in a given projection. We impose that, when observed in a certain projection, a galaxy must receive at least a score of 4 (out of a maximum possible score of 12 = 2 (for sharp shells) x 3 (judges) x 2 (contrast levels)) in order for us to mark the galaxy as having shells in that projection. Using this condition, we find that $\sim$50 -- 60\% of the shell galaxies in our sample have shell structures visible in 2/3 projections. About $40\%$ of the galaxies have shells visible in all 3 projections, and a handful of galaxies $\lesssim 5\%$ have shells visible in only one projection. We then compare the results obtained by imposing score cuts per projection with the summed up scores from Figure~\ref{fig:scoreDist}. The rough boundaries separating galaxies by score depending on whether they have shells visible in 1/2/3 projections are marked with grey bands. These bounds agree well with a simple prediction if the judges gave exactly the same scores and we ignore score deviations between the two contrast levels. For example, a galaxy with shells in one projection would have scores ranging roughly between $(1,0,0)$ from each judge leading to a fractional score of $1/6$ to $(2,0,0)$ from each judge (fractional score of $2/6$). Similarly, a galaxy with visible shells in 2 projections, would have scores ranging from $(1,1,0)$ to as high as $(2,2,0)$, i.e. fractional scores between $2/6$ and $4/6$. In theory, galaxies with visible shells in all 3 dimensions could have scores as low as $(1,1,1)$, but we find that this is very uncommon -- most often, shells visible in all 3 dimensions are very sharp and well defined, corresponding to galaxies with high fractional scores ($\gtrsim 4/6$).

In the ideal case, Type I shells are invisible if the line of sight is perfectly aligned with the symmetry axis of the shells, yet they are visible in the other two projections.
Indeed, our score distribution, as well as detailed studies of the trajectories of individual shell-forming progenitors, show that almost all shells are visible in at least 2 out of 3 projections. 
The shells are best observed in those projections in which the trajectory of the progenitor is very radial (e.g., the top panels of Figure~\ref{fig:projectionsColormaps}). In those projections in which the satellite spirals in towards the host, the shells are less visible (e.g., right-most middle panel of Figure~\ref{fig:projectionsColormaps}). There are a handful of galaxies for which we can see much sharper shells in one of the projections, but we cannot completely rule out shells in a second projection (see last row of Figure~\ref{fig:projectionsColormaps}). In these cases, the progenitor has a nearly-perfectly radial orbit in one projection, but nearby independent substructure or the exact details of the trajectory could make the shells slightly less visible in the second projection (bottom middle panel).

According to Figure \ref{fig:scoreDist}, most shells are visible in $2/3$ projections, with about $\sim$40\% of the galaxies having shells visible in all projections. Making the simplifying assumption that half of the shells are visible in $2/3$ projections and the other half are visible in all 3 projections, we find that the observed shell fractions should be about $5/6$ lower than the fractions obtained in Figure~\ref{fig:massDistofShells}. That is because, unlike the simulation results for which we used all 3 available projections, observers would only detect $2/3$ of the systems with shells in 2 projections and $100\%$ of those with shells in all projections, i.e. $1/2 \times 2/3 + 1/2 \times 1 = 5/6$. Using the exact projection distributions for our sample of galaxies, we find that observed fractions should be $f_{\mathrm{obs}} \sim 0.8 f_{\mathrm{sim}}$. In particular, the resulting observed fraction of shells in ellipticals with $\mathrm{M}_{\mathrm{200crit}} > 6 \times 10^{12}\, \mathrm{M}_\odot$ at $z=0$ would drop to $14\% \pm 3\%$ (compared to $18\% \pm 3\%$ in Figure~\ref{fig:massDistofShells}).

\section{Logistic Regression for Classifying Shell and Non-Shell-Forming Progenitors}
\label{sec:logisticRegression}
In Section~\ref{sec:progenitors}, we find the zero-order recipe for merger events that produce redshift $z=0$ shells.
Figure~\ref{fig:threeDim} investigates the region of the parameter space ($v_r/|v|$, $\mu_{\mathrm{stars}}$, $t_{\mathrm{acc}}$) preferred by those satellites that are successful at forming $z=0$ shells.
In order to disentangle the correlation between radial velocity ratio and stellar mass ratio of the satellites, we use logistic regression, also known as logit regression \citep{Cox1958}.
For this method, the dependent variable is categorical -- we mark with "1" shell-forming progenitors and with "0" any other satellites unable to form $z=0$ shells. 
While linear regression is the method of choice for regression problems, in this case we want to classify shell-forming/non-shell-forming satellites, and for this we use the equivalent method used for classification problems, i.e., we fit the logistic function:
\begin{eqnarray}
F(X) = \frac{1}{1 + e^{-(\alpha + \beta X)}}
\end{eqnarray}
Here, $F(X)$ can be interpreted as the probability that the dependent variable is equal to a success (i.e., shell-forming progenitors) vs. a failure (i.e., satellite doesn't form shells), while $X$ probes the two-dimensional space ($v_r/|v|$ vs. $\mu_{\mathrm{stars}}$). 
Since in our sample we have significantly more satellites that fail to form shells than shell-forming ones, we use balanced logistic regression.
The best-fit line separating shell-forming satellites from all other satellites is shown with a red continuous line in Figure~\ref{fig:threeDim}; this line corresponds to $P(\mathrm{class})=0.5$, i.e. we are equally likely to find shell-forming or non-shell-forming satellites. 
The dashed lines in the figure show $1\sigma$ bounds for this line.

By using logistic regression, we assume that the decision boundary separating shell-forming from non-shell-forming satellites is a simple linear function depending on $v_r/|v|$ and $\mu_{\mathrm{stars}}$. In order to avoid making this assumption, we run a discriminative classifier -- Support Vector Machine (SVM) \citep{Cortes&Vapnik1995}, applying the kernel trick to maximum-margin hyperplanes \citep{Boser1992}. This allows us to test different non-linear kernels (e.g., polynomials and Gaussian radial basis function). 
While the limited size of our sample of shell-forming progenitors doesn't allow us to rule out non-linear correlations in the ($v_r/|v|$, $\mu_{\mathrm{stars}}$) space, our initial tests suggest that the data sample presented in this paper prefers the red demarcation line shown in Figure~\ref{fig:threeDim}.

\section{Resolution Limits}
\label{sec:resolution}

\begin{figure*}
\vspace{-0em}\includegraphics[width=\textwidth]{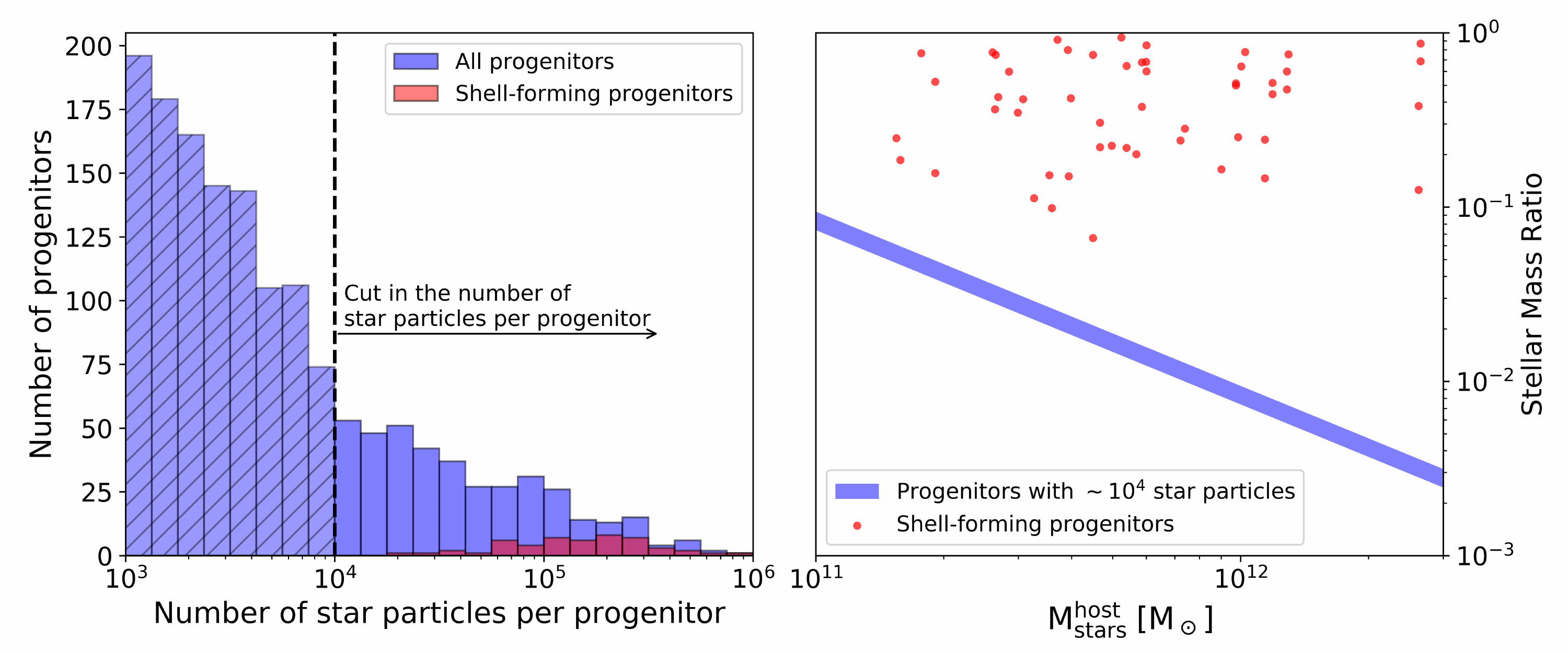}
\caption{\textit{Left:} In this study we consider satellites with at least $10^4$ star particles. The histogram above shows the distribution of the number of star particles per satellite for all the satellites accreted by shell galaxies in our sample. We highlight in red those satellites we identify as producing $z=0$ shells. The fraction of progenitors that are successful at forming shells drops rapidly for satellites with less than $\sim 5\times 10^4 $ star particles. We note that we can still clearly distinguish shells composed of as few as $2\times10^4$ star particles, as exemplified in the bottom row of Figure~\ref{fig:satelliteofSatellite}. 
\textit{Right:} The red scattered points show the distribution of stellar mass ratios of shell-forming progenitors as a function of the host stellar mass. The blue band corresponds to the typical minimum stellar mass ratios that we can probe when we restrict our study only to satellites with more than $10^4$ star particles.
Consequently, we claim that we do not miss any shell-forming progenitors for the range of massive host galaxies we include in our sample, and we are not resolution limited in observing shells composed of $>10^4$ star particles.}
\label{fig:resolution}
\end{figure*}

For the current study, we used the highest-resolution run in Illustris. As shown by \cite{Rodriguez-Gomezetal2015}, galaxy merger rates and mass ratios in Illustris measured at $t_{\mathrm{acc}}$ are well converged. 
The gravitational softening length for star particles in Illustris ($0.7$ kpc) is less than $\sim$10\% of the half-light radius for galaxies in our sample.
Furthermore, shells are situated at large galactocentric distances (outer shells extending to $\gtrsim 50-100$ kpc), in the low surface brightness regions of the stellar halos, and their self-gravity is essentially negligible. 

By focusing on massive elliptical galaxies in Illustris ($\mathrm{M}_{200\mathrm{crit}} > 6 \times 10^{12} \, \mathrm{M}_\odot$ and $\mathrm{M}_{\mathrm{stars,tot}} > 10^{11} \, \mathrm{M}_\odot$), our sample includes galaxies with at least $2.5 \times 10^5$ star particles each. 
In the current study, we only consider accreted satellites with at least $10^4$ star particles (corresponding to a stellar mass $\mathrm{M}_{\mathrm{stars,tot}} \gtrsim 9 \times 10^9 \,\mathrm{M}_\odot$). 
This allows us to resolve stellar shells composed of only a few percent of the stars in the main host.
The histogram in Figure~\ref{fig:resolution} shows the distribution of the number of star particles per satellite for all the satellites accreted by the 39 shell galaxies in our sample, selected using the first visual identification step (\S\ref{subsubsec:step1Visual}). The red histogram corresponds to those satellites we identify as producing shells at $z=0$ using the second identification step based on stellar history catalogs (\S\ref{subsubsec:step2Confirm}). The fraction of progenitors that are successful at forming shells drops rapidly for satellites with less than $\sim$5$\times 10^4 $ star particles. 

In the right panel of Figure~\ref{fig:resolution}, we compare the distribution of stellar mass ratios of shell-forming progenitors (in red) with the minimum stellar mass ratios that we can probe for satellites having at least $10^4$ star particles (blue band). Even for the smallest host galaxies in our sample, our cut at $10^4$ star particles would have allowed us to probe lower merger mass ratios than the ones we found for Illustris shell galaxies.

Moreover, we identify the shell-forming progenitor with the smallest number of star particles ($19241$) and $\mu_{\mathrm{stars}} = 0.066$ to correspond to a special case of a satellite of a satellite (see discussion in \S\ref{subsec:satofsat}). Based on our results for the order-zero recipe for forming shells presented in Figure~\ref{fig:threeDim}, satellites with such small stellar mass ratios have a small probability to form shells in massive $z=0$ hosts.
We establish in \S\ref{subsec:satofsat} that this satellite is a special case and that it forms shells by tagging along the orbit of its more massive companion. 
As a result, we consider that satellites with such low $\mu_{\mathrm{stars}}$ (and correspondingly low $\mathrm{N}_{\mathrm{star\,particles}}$) are unlikely to form shells for the host galaxies in our sample.
Nonetheless, while this type of shell-forming progenitor is very uncommon in our sample of satellites, we note that we can clearly distinguish the shells it forms despite the rather low number of star particles ($19241$) in their composition (see bottom row of Figure~\ref{fig:satelliteofSatellite}). Consequently, we claim that we do not miss any shell-forming progenitors by applying a cut of $>10^4$ star particles for the satellites we investigate in the second step of our method (\S\ref{subsubsec:step2Confirm}), and moreover, we are not resolution limited in observing shells composed of $>10^4$ star particles.

\section{Stellar Mass Completeness}
\label{sec:complete}

\begin{figure*}
\vspace{-0em}\includegraphics[width=0.7\textwidth]{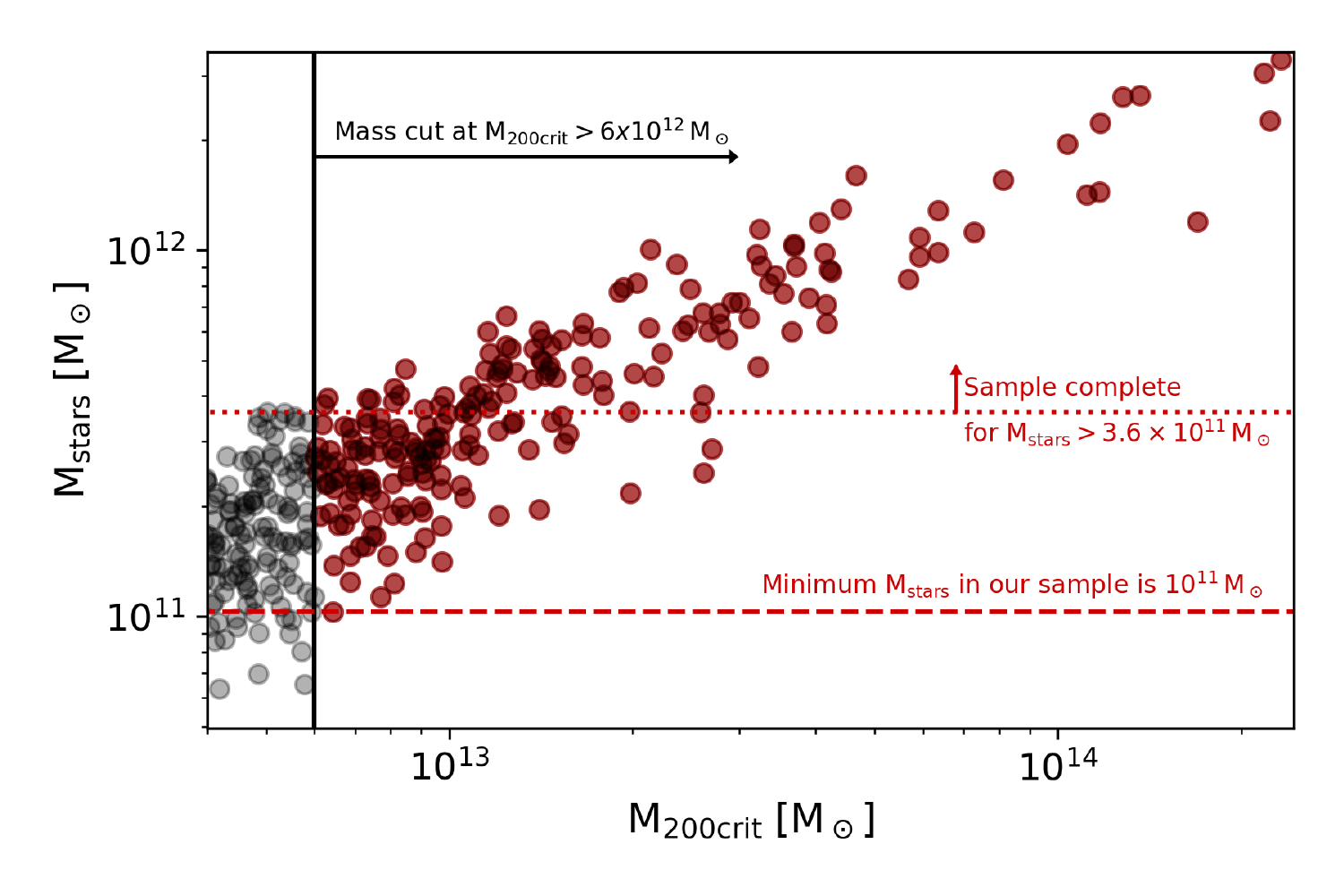}
\caption{Mass distribution of Illustris galaxies (shown in grey), with galaxies included in the current study ($\mathrm{M}_{\mathrm{200crit}} \geq 6 \times 10^{12} \,\mathrm{M}_\odot$) highlighted in red. The minimum stellar mass of galaxies in our sample is $10^{11} \,\mathrm{M}_\odot$, and our sample is complete for $\mathrm{M}_{\mathrm{stars}} > 3.6 \times 10^{11} \,\mathrm{M}_\odot$.}
\label{fig:completeness}
\end{figure*}

In Figure~\ref{fig:completeness}, we show the total and stellar mass distribution of galaxies included in our sample. The overall Illustris galaxy population is shown in grey, while our galaxy sample is highlighted in red. We select the galaxies based on a mass cut: $\mathrm{M}_{\mathrm{200crit}} \geq 6 \times 10^{12} \,\mathrm{M}_\odot$. Our sample is stellar mass complete above 
$\mathrm{M}_{\mathrm{stars}} > 3.6 \times 10^{11} \,\mathrm{M}_\odot$, while 
the smallest galaxies in our sample have stellar masses $\sim 10^{11} \,\mathrm{M}_\odot$.

\bibliographystyle{mnras}
\bibliography{IllustrisShellGalaxies}

\label{lastpage}

\end{document}